\documentclass[twocolumn]{aa}
\usepackage[hidelinks,colorlinks=false]{hyperref}
\usepackage{graphicx}
\usepackage{txfonts}
\begin{document}

\title {New insight into the hard X-ray emission influenced by  the type-\uppercase\expandafter{\romannumeral1} bursts observed by Insight-HXMT during outburst of 4U 1636--536}
\titlerunning{type-\uppercase\expandafter{\romannumeral1} bursts of 4U 1636--536}

\author
{
J. Q. Peng\inst{1,2}\thanks{E-mail: pengjq@ihep.ac.cn} \and
S. Zhang\inst{1}\thanks{E-mail: szhang@ihep.ac.cn}\and  Y. P. Chen\inst{1}\thanks{E-mail: chenyp@ihep.ac.cn}\and L. D. Kong\inst{3} \and P. J. Wang\inst{3} \and S. N. Zhang\inst{1,2} \and   Q. C. Shui\inst{1,2} \and L. Ji\inst{4}\and G. B. Zhang\inst{2,5,6,7}\and Z. Yan\inst{2,5,6,7}\and
L. Tao\inst{1}\and J. L. Qu\inst{1,2}\and   M. Y. Ge\inst{1}\and Z. L. Yu\inst{1}\and J. Li\inst{8,9}\and Z. Chang\inst{1}\and Z. S. Li\inst{10}\and P. Zhang\inst{11,12}\and Y. X. Xiao\inst{1,2}\and S. J. Zhao\inst{1,2}}

\institute{Key Laboratory of Particle Astrophysics, Institute of High Energy Physics, Chinese Academy of Sciences, Beijing 100049, China
\and University of Chinese Academy of Sciences, Chinese Academy of Sciences, Beijing 100049, China
\and  Institute f{\"u}r Astronomie und Astrophysik, Kepler Center for Astro and Particle Physics, Eberhard Karls, Universit{\"a}t, Sand 1, D-72076
T{\"u}bingen, Germany \and School of Physics and Astronomy, Sun Yat-Sen University, Zhuhai, 519082, China \and Yunnan Observatories, Chinese Academy of Sciences, Kunming 650216, China\and Key Laboratory for the Structure and Evolution Celestial Objects, Chinese Academy of Sciences, Kunming 650216,  China \and Center for Astronomical Mega-Science, Chinese Academy of Sciences, Beijing 100012, China\and CAS Key Laboratory for Research in Galaxies and Cosmology, Department of Astronomy,
University of Science and Technology of China, Hefei 230026, China\and School of Astronomy and Space Science, University of Science and Technology of China, Hefei 230026, China \and Key Laboratory of Stars and Interstellar Medium, Xiangtan University, Xiangtan 411105, Hunan, China \and College of Science, China Three Gorges University, Yichang 443002, China \and Center for Astronomy and Space Sciences, China Three Gorges University, Yichang 443002, China }

\authorrunning{J. Q. Peng et al.}
\date{Received ; accepted }

 
  \abstract
   {Thermonuclear bursts, also known as type-I X-ray bursts, result from unstable nuclear burning of H/He accreted to the surface of neutron stars, lasting from tens of seconds to hundreds of seconds. Thermonuclear bursts have an important impact on accretion environments around the neutron stars, such as their disks and corona, and are therefore a subject of extensive research. 
   Thermonuclear bursts can be used as probes to gain a deeper understanding of the properties of their disks and corona.}
   {By analyzing the data from Insight-HXMT and NICER, we can determine the evolution of the significance of the hard shortage in 4U 1636--536 with its spectral state, as well as the evolution of the fraction of deficit with energy. Additionally, we investigate the possible geometry and evolution of the corona in 4U 1636--536 by combining our findings with the results of spectral analysis.}
   {Extracting the light curves from the Insight-HXMT LE
, ME, and HE data and subtracting their pre-burst emission allows us to estimate the significance of the hard shortages during the bursts. By fitting the spectra, the correlation between the persistent spectral parameters and the significance of the hard shortages can be established. The bursts are then grouped according to the spectral state in which they occurred, and the significance of the hard shortages is estimated. These in turn help to investigate the evolution of the fraction of deficit with energy.}
   {We find that during the soft state, the significance of possible hard X-ray shortage in bursts is almost zero. However, in the hard state, some bursts exhibit significant shortages (>3 $\sigma$), while others do not. We attempt to establish a correlation between the significance of the hard X-ray shortage and the spectral parameters, but the data quality and the limited number of bursts prevent us from finding a strong correlation.
   For bursts with insignificant shortages in the soft state, their fraction of the deficit remains small. 
   However, in the hard state, the fraction of deficit for all bursts increases with energy, regardless of the significance of the shortage of individual bursts.
   For bursts during the hard state, we investigate the evolution of the fraction of deficit during the bursts by stacking the peaks and decays of the bursts, respectively, and find that as the flux of the bursts decreases, the energy corresponding to the maximum of the fraction of deficit becomes progressively higher. }
{We explore the possible geometry and evolution of the corona clued by the evolution of the fraction of deficit, which is obtained from the spectral and temporal analysis.}
  

   \maketitle
%

\section{Introduction}
A low-mass X-ray binary (LMXRB) consists of a companion and a compact star, such as a black hole or neutron star. 
LMXRBs can be classified into persistent or transient sources \citep{2016Tetarenko,2018Sreehari}. 
The transient sources show occasionally sudden increases in X-ray flux through accretion, referred to as outbursts.
Neutron star LMXRBs can be classified into z-sources and atoll sources. Z-sources are significantly brighter than atoll sources, but more type-\uppercase\expandafter{\romannumeral1} bursts can be seen in atoll sources \citep{1993Lewin,2008Galloway}. Interestingly, z-sources and atoll sources can be transformed between each other \citep{2009Lin}.
With the evolution of outbursts, we can observe changes in their spectral state on the Color-Color Diagram (CCD). The trajectory of an atoll source on the CCD during an outburst typically follows a 'C' or 'U' shape. The spectral state of atoll sources can be classified into two states: the hard state and the soft state \citep{1989Hasinger,1989Schulz}. During the hard state, the spectral energy distribution of LMXRBs is dominated by hard X-ray emission, typically with a power-law shape. This emission is produced by the Comptonization of soft photons in the hot corona surrounding the accretion disk. In contrast, during the soft state, the spectrum is dominated by the thermal emission of the accretion disk, which emits a blackbody-like spectrum.

Type-I bursts are caused by unstable burning of the accreted hydrogen/helium on the surface of a neutron star \citep{1975Hansen,1976Belian}.
The flux of the bursts can be several times higher than the pre-burst persistent emission. The flux rise time of the classical X-ray bursts is between 1 and 10 seconds, while their duration can span from tens to hundreds of seconds.  The light curves show the characteristics of “fast rise and slow decay” \citep{1993Lewin,2004Cumming,2006Strohmayer}.

The influence of these bursts on the disk and corona is of great importance for studying the structure and evolution of these components.
\cite{2018Fragile,2020Fragile} has demonstrated through simulation that the presence of Poynting–Robertson (PR) drag causes the inner edge of the disk to retreat from the neutron-star surface toward larger radii and then recover on the timescale of the burst. 
\cite{2018Keek} reported a soft excess in Aql X--1 and provided an interpretation by introducing a reprocessing of the strongly photoionized disk and an enhancement of the pre-burst persistent flux, probably due to PR drag.
\cite{2023Lu}  also reported the enhanced persistent emission in 4U 1730--22 due to PR drag or the reflection from the accretion disk with the observation of NICER.
\cite{2019Bult} and \cite{2020Buisson} also reported soft excesses in SAX J1808.4--3658 and Swift J1858.6--0814 with NICER, respectively.  \cite{2003Maccarone} showed a hard X-ray shortage in an X-ray burst using RXTE/HXETE, with a significance of 2 $\sigma$. 
\cite{2012Chen} stacked the light curves of several X-ray bursts of IGR J17473--2721 with the observation of RXTE/PCA, and found the hard X-ray shortage with high confidence for the first time. 
Later on, a series of hard X-ray shortages were reported in Aql X--1 \citep{2013Chen}, 4U 1636--536 \citep{2013Ji,2019Chen},  KS 1731--260 \citep{2014J}, 4U 1705--44 \citep{2014J}, GS 1826--238 \citep{2014Ji} and 4U 1728--34 \citep{2017Kajava}, MAXI J1816--195 \citep{2022Chen} based on RXTE/PCA, INTEGRAL and Insight-HXMT observations, respectively.  \cite{2022Chen} gave the evolution of the deficit of fraction with energy in the millisecond accreting system MAXI J1816--195. They notice a clue that the fraction of the deficit drops at $\sim$  65--85 keV, which is probably attributed to hard X-rays from the accretion column of the magnetic pole, where the emissions are less influenced by the burst.

4U 1636--536 is a Neutron Star low-mass X-ray binary (LMXRB) with a companion star of mass $\sim$ 0.4 $M_{\odot}$ discovered by the 8th Orbiting Solar Observatory (OSO--8) \citep{2002Giles}. The track of 4U 1636--536 on the CCD is C-shape or U-shape, behaving as a standard atoll source \citep{1989Hasinger,1989Schulz,2011Zhang}. The orbit of 4U 1636--536 is about 3.8 hours \citep{1990van} and the spin period is 581 Hz \citep{1998Strohmayer}. \cite{2006Galloway} used the type-I bursts with photospheric radius expansion (PRE) to estimate a distance of $\sim$  6 kpc.
Since the discovery of 4U 1636--536, more than 700 bursts have been observed, including some PRE bursts, single-peaked bursts, and multi-peaked bursts \citep{2008Maurer,2009Zhang,2020Galloway}.
In the study of multi-peaked type-I X-ray bursts in the Neutron Star LMXB 4U 1636--536 with RXTE,  the first quadruple-peaked burst has been reported. \citep{2021Li}.
\cite{2013Ji} detect the  hard X-ray shortage (30--50 keV) in 4U 1636--536 by stacking 114 RXTE bursts. The hard X-ray shortage lags that of the soft X-rays by 2.4$\pm$1.5 s. The time lag was regarded as a timescale for cooling and recovering the corona. \cite{2019Chen} reported Insight-HXMT observation of the high energy shortage induced by a single burst, with a significance of 6.2 $\sigma$. \cite{2022Zhao} found enhanced persistent emissions possibly due to the PR drag and disk reflections during the bursts with NICER observations.
\cite{2022Guver} also revealed soft excess in 4U 1636--536, which is attributed to the increased mass accretion rate onto the neutron star due to PR drag. Also, they found evidence of corona cooling by stacking 7 NuSTAR bursts.

In this paper, we analyze the burst data of 4U 1636--536 collected by Insight-HXMT and NICER and investigate in detail the burst influence on its corona in the context of outburst evolution. In Section \ref{obser}, we describe the observations and data reduction. The detailed results are presented in Section \ref{result}. The results are discussed and the conclusions are presented in Section \ref{dis}.


\section{Observations and Data reduction}
\label{obser}

\subsection{Insight-HXMT}
Insight-HXMT is the first Chinese X-ray astronomy satellite, which was successfully launched on 2017 June 15 \citep{2014Zhang, 2018Zhang, 2020Zhang}. It carries three scientific payloads: the low energy X-ray telescope (\textrm{LE}, SCD detector, 1--15 keV, 384 $\rm cm^{2}$,
\citealt{2020Chen}), the medium energy X-ray telescope (\textnormal{ME}, Si-PIN detector, 5--35 keV, 952 $\rm cm^{2}$, \citealt{2020Cao} ), and the high energy X-ray telescope (HE, phoswich NaI(CsI), 20--250 keV, 5100 $\rm cm^{2}$, \citealt{2020Liu}).

Insight-HXMT has observed 4U 1636--536 from 2018 to 2022, with a total exposure time of 1050 ks.
We extract the data from all three payloads using the Insight-HXMT Data Analysis software {\tt{HXMTDAS v2.05}}. 
The data are filtered with the criteria recommended of the Insight-HXMT Data Reduction Guide {\tt v2.05.}\footnote[1]{{http://hxmtweb.ihep.ac.cn/SoftDoc/648.jhtml}}

We extract the total light curves with a time-bin of 1 second (including instrument background and source counts) for LE (1--10 keV) and ME (6--30 keV), and a time-bin of 4 seconds for HE (40-70 keV). Then extract the light curves during the bursts with the peak of the ME light curves of the bursts as the time zero (Figure \ref{lc}). {\tt Xspec v12.12.1}\footnote[2]{{https://heasarc.gsfc.nasa.gov/docs/xanadu/xspec/index.html}} is used to fit the spectra and the energy bands are chosen as, LE 2--8 keV, ME 8--28 keV and HE 28--100 keV.
One percent systematic error is added to data, and errors are estimated via Markov Chain Monte-Carlo (MCMC) chains with a length of 20000.

\subsection{NICER}
The Neutron Star Interior Composition Explorer (NICER) is an International Space Station (ISS) payload, which was launched by the Space X Falcon 9 rocket on 3 June 2017. And NICER has a large effective area and high temporal resolution in the soft X-ray band (0.2--12 keV).
NICER observed 4U1636--536 in 2018 and had four observations almost simultaneous with those of Insight-HXMT. 
NICER data are reduced using the standard pipeline tool {\tt NICERl2\footnote[3]{https://heasarc.gsfc.nasa.gov/lheasoft/ftools/headas/nicerl2.html}}. The background is calculated by "{\tt nibackgen3C50\footnote[4]{https://heasarc.gsfc.nasa.gov/docs/nicer/analysis\_{}threads/background/}}" tool provided by the NICER team. We group each spectrum as having at least 50 counts in each channel, and for the fitting of the spectrum we choose an energy range of 0.5--10 keV.

\section{Result}
\label{result}

\subsection{Color-color diagram and Insight-HXMT light curves}
\label{ccd}

\begin{figure}
	\centering
	\includegraphics[angle=0,scale=0.38]{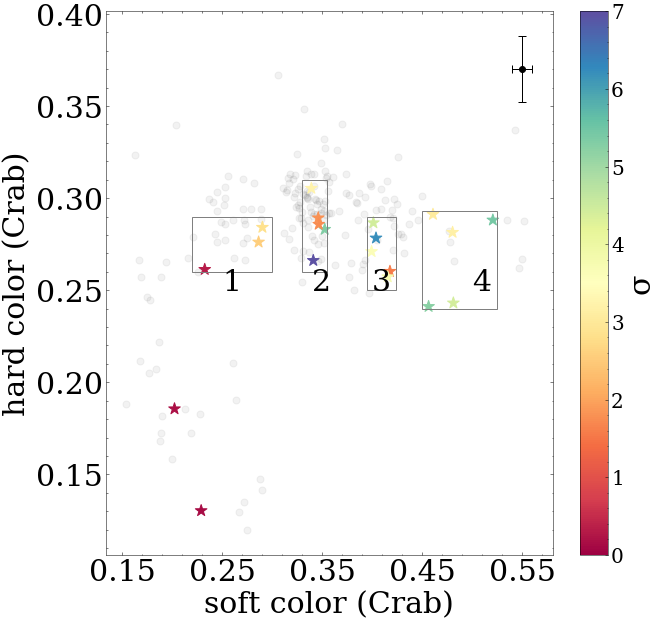}
	\caption{The Insight-HXMT CCD of 4U 1636--536, where soft-color is defined as the ratio of LE 3--6 keV to 1--3 keV count rates, and the hard-color is defined as the ratio of ME 15--20 keV to 10--15 keV count rates. Correction is made by dividing by the count rate of the Crab corresponding to the time and energy band. The time bin of the points is 4 hours. The stars are type-I bursts used in the paper. The average error bar is shown in the top right corner. The Color bar shows the levels of significance of hard X-ray shortage during the bursts. The bursts in the hard state are classified into four groups based on their positions on the CCD.
}
	\label{CCD}
\end{figure}

\begin{figure}
	\centering
\begin{minipage}{0.45\textwidth}
  \includegraphics[angle=0,scale=0.07]{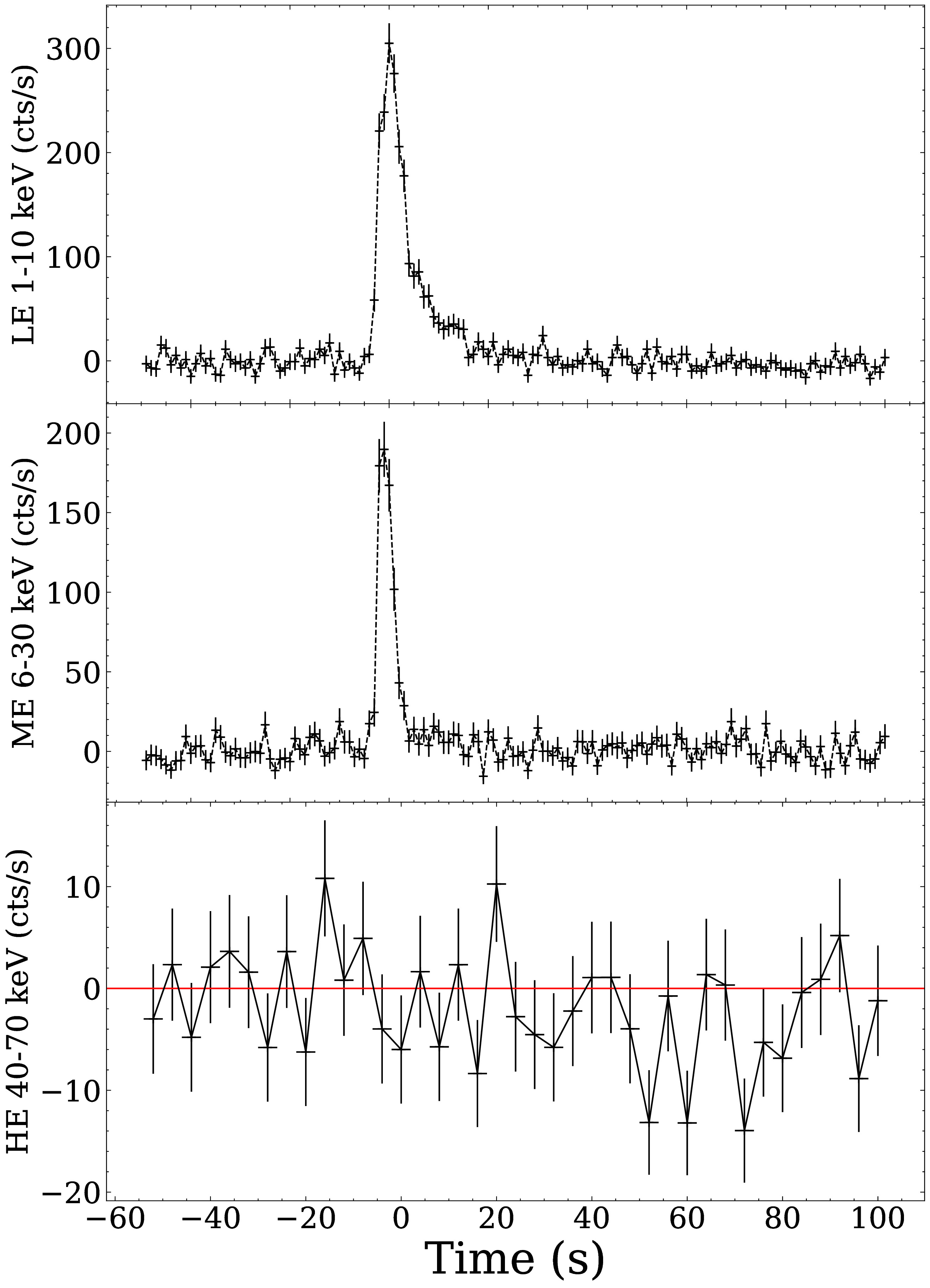} \\
  \includegraphics[angle=0,scale=0.07]{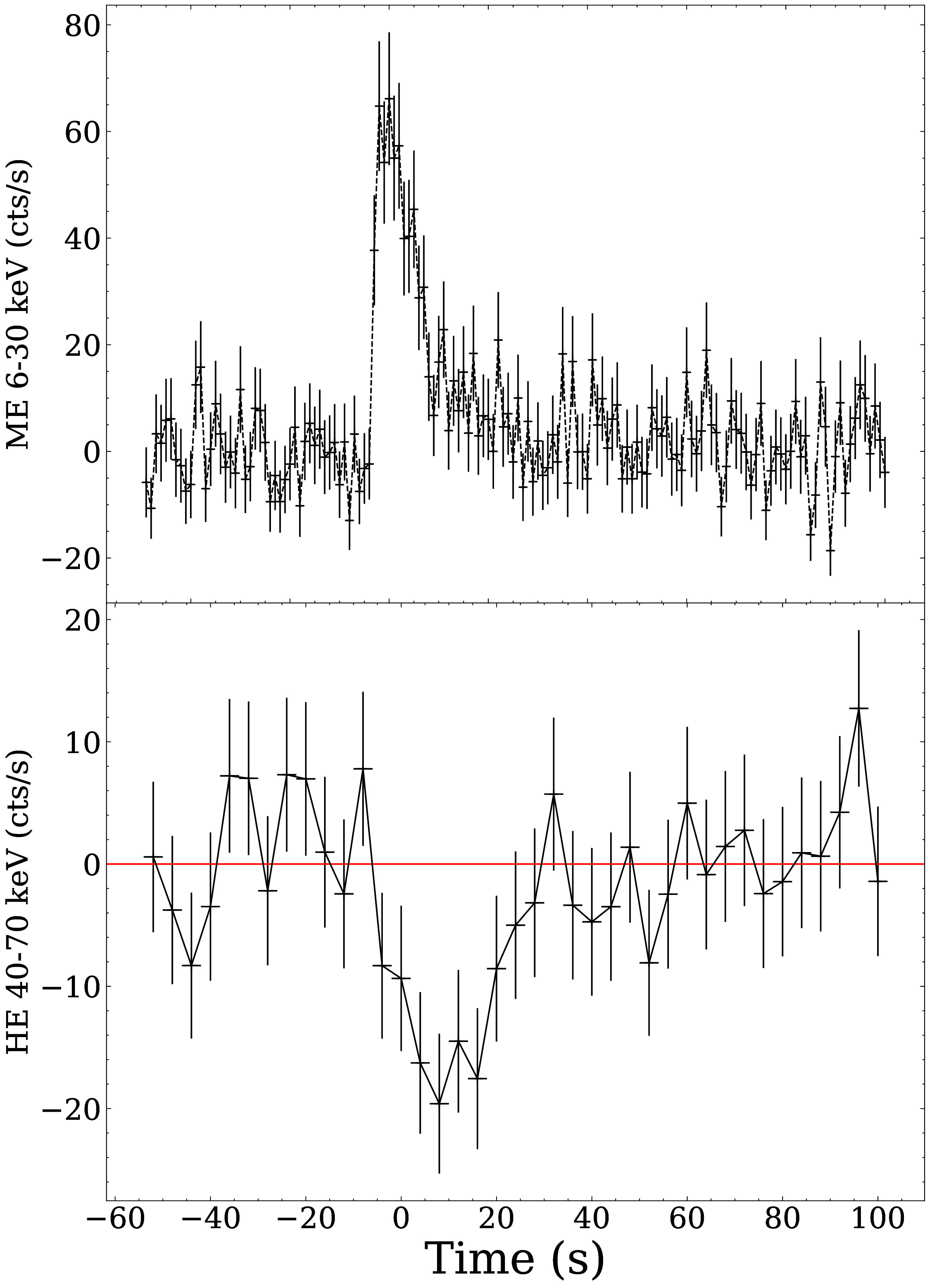}
  \caption{Two examples of Insight-HXMT light curves: top three panels for a burst in soft state, bottom two panels for a burst in hard state with the most significant hard X-ray shortage. The panels from top to bottom are the LE, ME, and HE light curves during the burst of 4U 1636--536 in 1--10 keV, 6--30 keV, and 40--70 keV. respectively.
  Time 0 s is the peak flux of the ME. For all bursts, we take the mean value of the pre-burst emission (both persistent emission and instrumental background) as the background and subtract it from the light curves of LE, ME and HE to obtain the net light curves, respectively.}
  \label{lc}

\end{minipage}
\end{figure}

\begin{table*}
\renewcommand\arraystretch{2}
\normalsize
	\begin{center}
		
		\caption{The bursts detected by Insight-HXMT  of 4U 1636--536.}
		\vspace{-0.3cm}
		\begin{tabular}{cccccc}
		\\ \hline
        ObsID & State & Time(MJD) & Significance($\sigma$) & Rise Time (s)&e-folding Time(s)
        \\ \hline
        P011465400401-20180215-01-01 & soft & 58164.72 & 0.19&3.0&9.5 \\ \hline
        P011465400501-20180217-01-01 & soft & 58166.05 & 0.11&2.5&4.0 \\ \hline
        P011465400901-20180226-01-01 & hard & 58175.85 & 0.31&4.0&7.5 \\ \hline
        P011465402101-20180403-01-01 & hard & 58211.18 & 1.57&6.5&28.7 \\ \hline
        P011465402701-20180630-01-01 & hard & 58299.43 & 3.24&6.7 &25.5 \\ \hline
        P011465402801-20180701-01-01 & hard & 58300.69 & 5.31&6.5&22.2 \\ \hline
        P011465403301-20180706-01-01 & hard & 58305.66 & 2.84 &2.0&19.7\\ \hline
        P011465403601-20180709-01-01 & hard & 58308.99 & 2.56 &3.0&7.6\\ \hline
        P030508400201-20210410-01-01 & hard & 59314.91 & 4.41&2.8&19.5 \\ \hline
        P030508400401-20210424-01-01 & hard & 59328.63 & 5.42&6.0&30.25 \\ \hline
        P030508400502-20210502-01-01 & hard & 59336.84 & 4.23&4.5&16.9 \\ \hline
        P030508400603-20210522-01-01 & hard & 59356.92 & 3.16&5.3&15.3 \\ \hline
        P030508400701-20210526-01-01 & hard & 59360.96 & 3.05&3.8&19.6 \\ \hline
        P030508400801-20210603-01-01 & hard & 59368.59 & 4.45&2.8&26.0 \\ \hline
        P030508401102-20210713-01-01 & hard & 59408.68 & 6.95&4.0&17.4 \\ \hline
        P040518100401-20220418-01-01 & hard & 59687.58 & 3.61 &4.0&16.52\\ \hline
        P040518100502-20220422-01-01 & hard & 59691.37 & 1.51&2.8&18.1 \\ \hline
        P040518100801-20220510-01-01 & hard & 59709.43 & 1.74 &4.0&18.0\\ \hline
        P040518100901-20220516-01-01 & hard & 59715.06 & 6.23 &3.3&16.5\\ \hline
        P040518101103-20220529-02-01 & hard & 59728.11 & 3.77&3.0&15.1 \\ \hline
        Observation of merged hard states & & & 15.64
        \\ \hline
        \label{obs} &     
		\end{tabular}
		
	\end{center}
	
\end{table*}

We subtract the background and get the net count rates of Insight-HXMT LE: 1--3 keV, 3--6 keV and ME: 10--25 keV, 15--20 keV, and then construct the Color-Color Diagram (CCD) of 4U 1636--536.
As shown in Figure \ref{CCD}, the soft color of the CCD is defined as the LE count rate ratio of 3--6 keV to 1--3 keV, while the hard color is the ME count rate ratio of 15--20 keV to 10--15 keV. We correct the CCD by taking into account the relative count rate ratio for Crab, using similar observation times for 4U 1636-536.
In Figure \ref{CCD}, the red dots represent type-I bursts used in the paper, and most of these bursts are in the hard state. 

For each burst, we extract the light curves of the Insight-HXMT 1--10 keV and 6--30 keV with a time-bin of 1 second for LE and ME, 40--70 keV with a time-bin of 4 seconds for HE. We use the time of the flux peak of each ME light curve with a time resolution of 1 second as a reference (time zero in Figure \ref{lc}). For all bursts, we subtract the pre-burst count rate, which includes both persistent emission and instrumental background, to have the net burst light curves for LE, ME, and HE, respectively. Shown in Table \ref{obs} are the significant levels of the hard X-ray shortages for each burst which are estimated as the ratio of the deficit area to the sum of all error bars during the deficit.
As shown in Figure \ref{lc} for a typical burst embedded in the soft state, the significance of the hard X-ray shortage is essentially zero in the 40--70 keV energy range. However, for bursts in the hard state, the hard X-ray shortage is quite significant, indicating a shortage of hard X-rays induced by bursts. The hard X-ray shortage detected in a single burst can be as large as around 7 $\sigma$, which turns out to be the most significant hard X-ray shortage detected so far from a single burst (see Figure \ref{lc}). Moreover, after combining the bursts of hard states, the significance of the hard shortage increases to 15.6 $\sigma$ (Table \ref{obs}), thus providing so far the most significant burst-induced hard X-ray shortage detected for 4U 1636--536.

\subsection{Evolution of the spectral parameters}
\label{spectral}
\begin{figure}
	\centering
	\includegraphics[angle=0,scale=0.07]{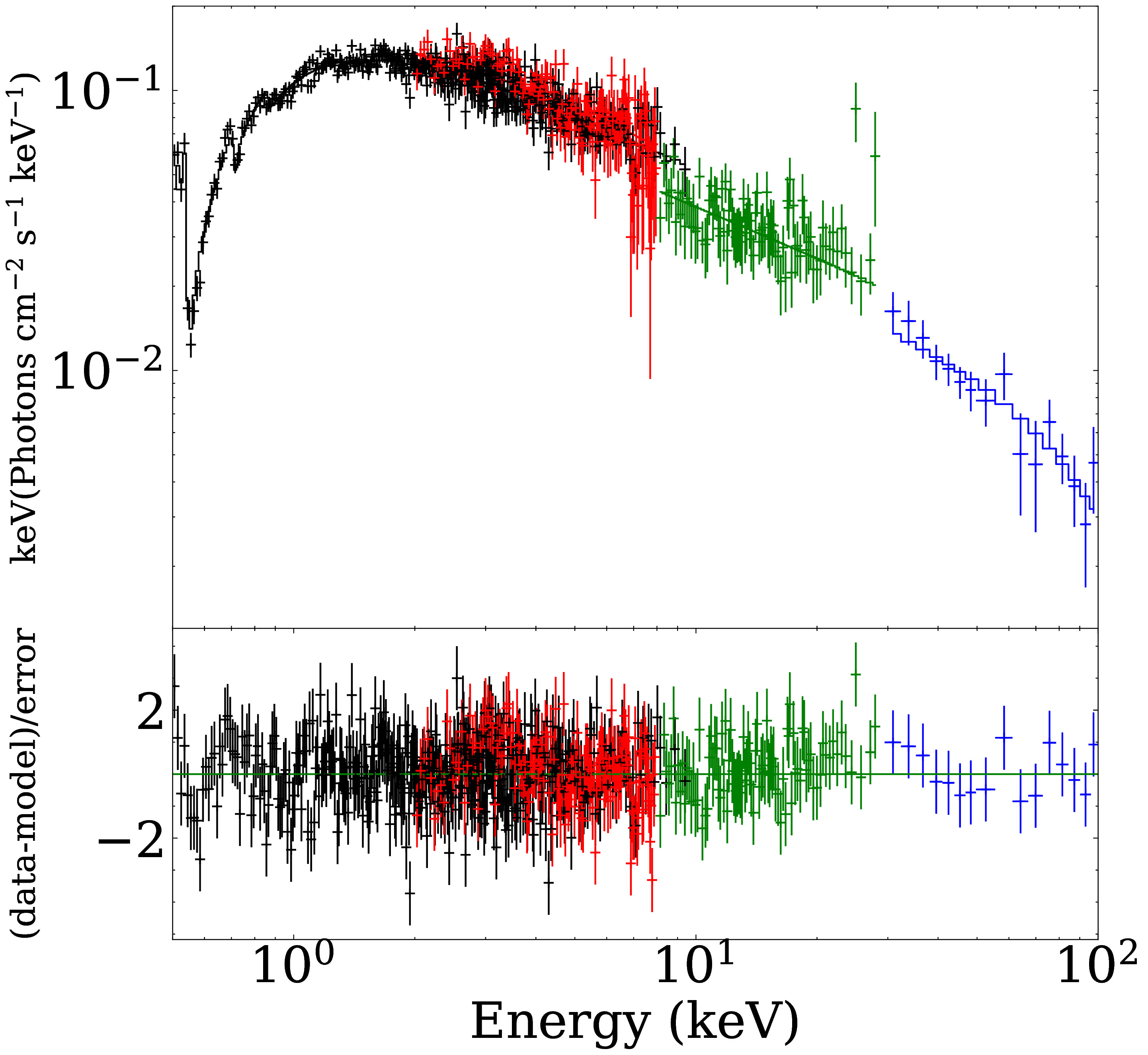}
	\caption{The simultaneous broadband spectrum of ObsID P011465402801 of 4U 1636--536 is observed from NICER (black), Insight-HXMT/LE (red), Insight-HXMT/ME (green) and Insight-HXMT/HE (blue). }
	
	\label{2801}
\end{figure}

\begin{figure*}
	\centering
	\includegraphics[angle=0,scale=0.07] {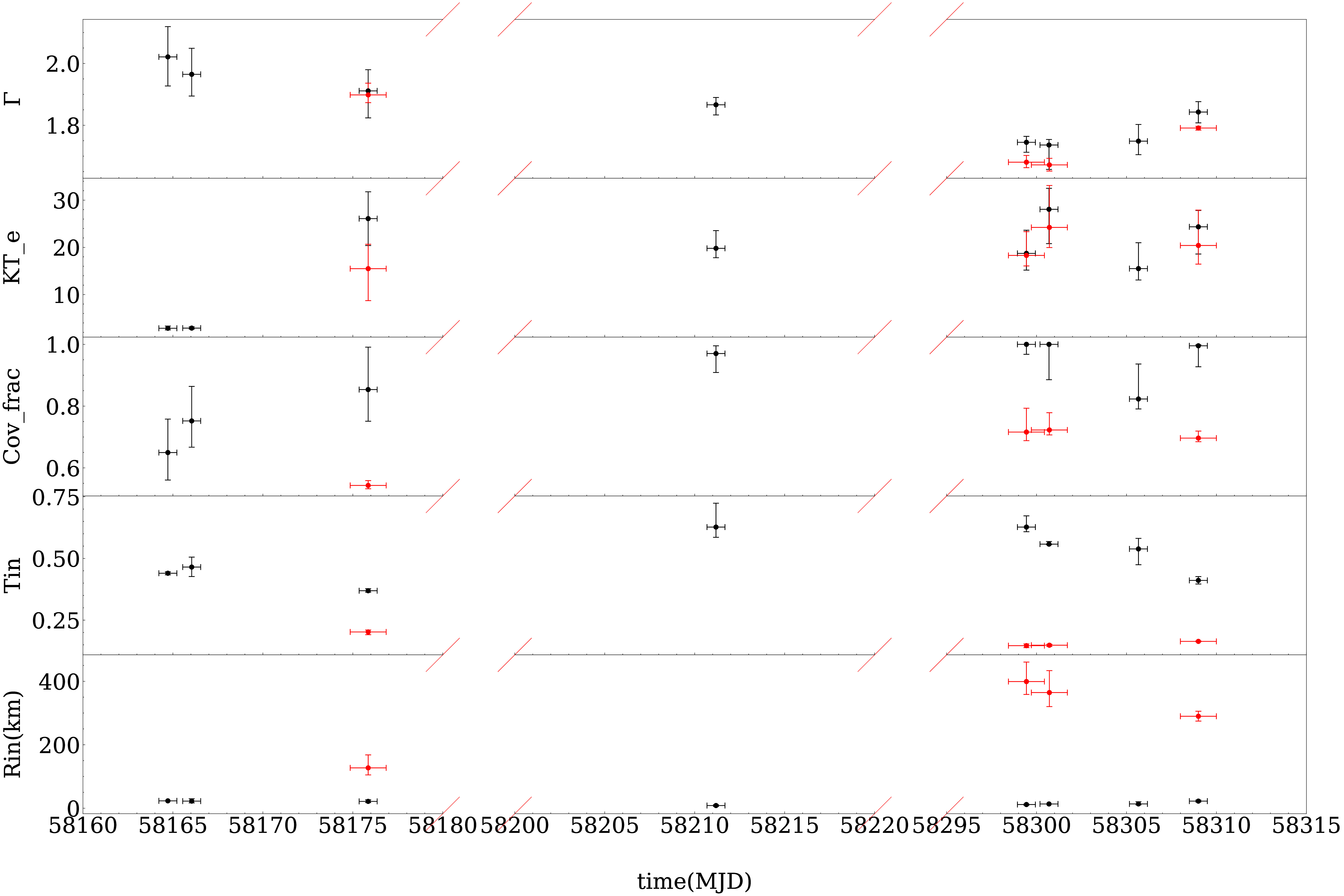}
	\caption{The evolution of parameters of the pre-burst continuum spectrum of 4U 1636-536 in 2018: $\Gamma$ are the low energy power-law photon index, $kT_{\rm e}$ the electron temperature, Cov\_frac the coverage factor, $T_{\rm in}$ the temperature of the inner disk, and $R_{\rm in}$ the inner radius of the disk. The black points are the parameters by fit only with the data of Insight-HXMT. The red points are the parameters by
	joint fitting of Insight-HXMT and NICER.}
	
	\label{com}
\end{figure*}

\begin{table*}

\renewcommand\arraystretch{2}
	\begin{center}
		
		\caption{Spectrum parameters of 4U1636--536 Insight-HXMT and NICER joint fitting.}
		\vspace{-0.3cm}
		\resizebox{\textwidth}{!}
		{
		\begin{tabular}{cccccccccccccccccc}
    \hline
    \hline
       Insight-HXMT & NICER & Time & $\Gamma$ & $kT_{\rm e}$ & Cov\_frac & $T_{\rm in}$ & $R_{\rm in}$ & $\sigma$ &Flux (40-70 keV) &Flux (10-100 keV) \\ObsID&ObsID&(MJD)& & keV& &keV& km& &$10^{-10}$~erg~s$^{-1}$~cm$^{-2}$ & $10^{-9}$~erg~s$^{-1}$~cm$^{-2}$& 
        \\
        \hline
        P011465400901 & 1050080146&58175 & $1.88^{+0.04}_{-0.03}$ & $15.5^{+6.77}_{-5.17}$ & $0.54^{+0.01}_{-0.02}$ & $0.21^{+0.01}_{-0.01}$ & $134.1^{+38.7}_{-21.1}$ & 0.301 & $1.03^{+0.04}_{-0.03}$ & $0.45^{+0.01}_{-0.01}$ \\
        \hline
        P011465402602 &1050080156 &58298 & $1.68^{+0.02}_{-0.02}$ & $18.3^{+2.23}_{-5.03}$ & $0.71^{+0.03}_{-0.08}$ & $0.14^{+0.01}_{-0.01}$ & $399.2^{+40.6}_{-61.4}$ & 3.240 & $7.39^{+0.12}_{-0.11}$ & $2.87^{+0.04}_{-0.04}$ \\ \hline
        P011465402801 &1050080157 &58300& $1.67^{+0.02}_{-0.02}$ & $24.2^{+14.27}_{-4.23}$ & $0.72^{+0.02}_{-0.06}$ & $0.15^{+0.01}_{-0.004}$ & $364.2^{+44.2}_{-69.2}$ & 5.319 & $8.38^{+0.17}_{-0.16}$ &$2.84^{+0.06}_{-0.06}$ \\ \hline
        P011465403601 &1050080157 &58308 & $1.79^{+0.01}_{-0.01}$ & $20.3^{+7.49}_{-4.94}$ & $0.69^{+0.02}_{-0.01}$ & $0.16^{+0.002}_{-0.002}$ & $289.8^{+15.2}_{-15.8}$ & 2.559 & $3.17^{+0.05}_{-0.06}$ & $1.29^{+0.03}_{-0.02}$ \\ \hline
			\label{joint}
		\end{tabular}
		}
	\end{center}
	
\end{table*}

\begin{figure*}
\centering

\begin{minipage}[t]{0.45\linewidth}
\centering
\includegraphics[angle=0,scale=0.07]{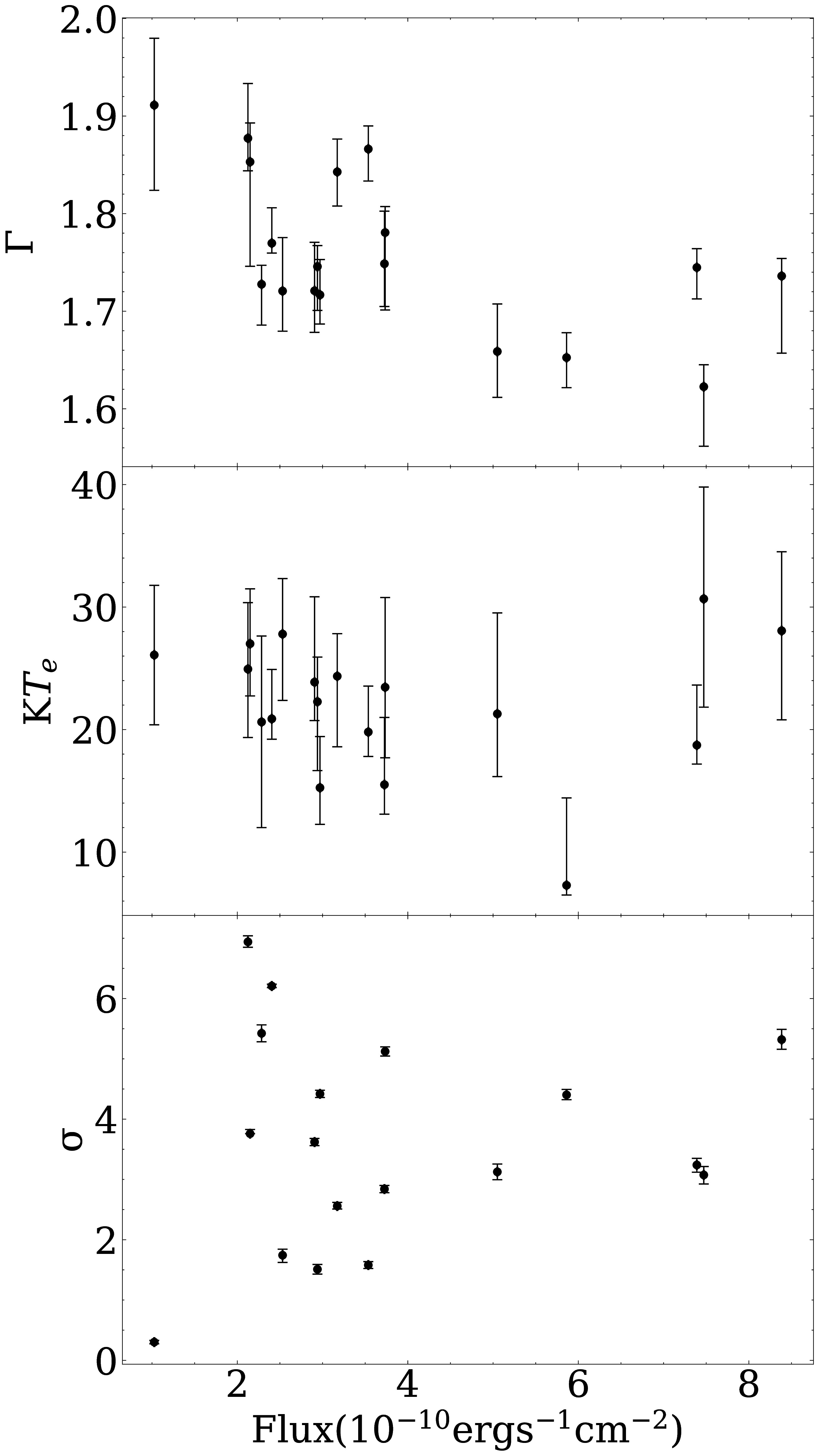}
\end{minipage}%
\hfill
\begin{minipage}[t]{0.45\linewidth}
\centering
\includegraphics[angle=0,scale=0.07]{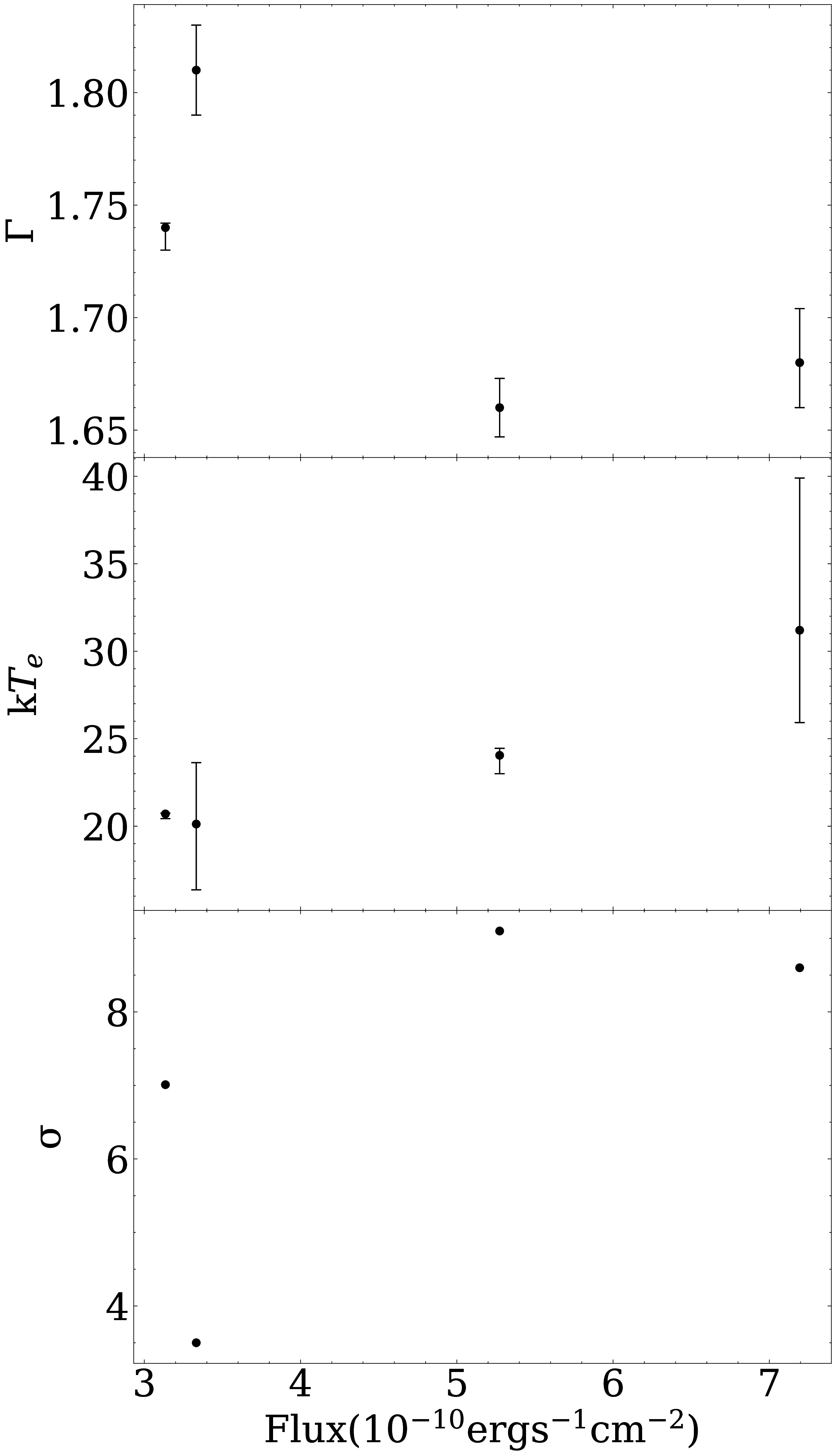}
\end{minipage}
\centering
\caption{The correlation of flux and persistent spectral parameters for 4U 1636--536. $\sigma$ the significance of hard X-ray shortage, $kT_{\rm e}$ the electron temperature, $\Gamma$ the photon index, and Flux the persistent emission flux of 40--70 keV. Left panel: Each point corresponds to the pre-burst continuum spectrum of the observation in which each burst of 4U 1636-536 is located. Right panel: We have stacked the bursts in the hard state, points from left to right corresponding to the stacking of the bursts in the four regions 1, 2, 3, and 4 of Figure \ref{CCD}, respectively. The spectral parameters are the result of the fit after combining the pre-burst continuum spectra of the observations where these bursts are located.}
\label{corr}
\end{figure*}

To investigate the relationship between the significance of hard X-ray shortage of each burst and the parameters of the continuum spectrum in which the bursts are located, we fit the pre-burst continuum spectrum.
We use TBABS component to account for the interstellar absorption \citep{2000Wilms}. Because of the calibration problem of LE, we ignore energies below 2 keV. Accordingly, the column density $ N\rm _H$ can not be well constrained and hence fixed at 0.3 $\times$ $10^{22}$ $\rm cm^{-2}$ \citep{2008Pandel}.
We multiply a CONSTANT to account for the calibration discrepancies between Insight-HXMT LE, ME, and HE (in this paper, we fix the constant of LE to 1). 
The DISKBB model is used to fit the multitemperature blackbody component of the accretion disk \citep{1984Mitsuda}.
At this step, an obvious Compton hump shows up in the residual [$\chi^2$/(d.o.f)=814/342=2.38].
So we convolve the THCOMP model with the DISKBB model to fit the thermally Comptonized continuum formed by the Comptonization of the thermal electrons \citep{2020Zdziarski} [$\chi^2$/(d.o.f)=327/339=0.96].
Therefore, our fitting model is CONSTANT*TBABS*(THCOMP*DISKBB). A typical joint spectral fit is shown in Figure \ref{2801}. 

As the flux of 4U 1636--536 is too low for the Insight-HXMT data to constrain the model parameters well, we take almost simultaneous observations of NICER and Insight-HXMT for joint fitting. 
The model used for fitting is the aforementioned one and the resulting spectral parameters are shown in Figure \ref{com}.
The red points are from the joint fit and the black are the result with Insight-HXMT data only.
We can see that the photon index $\Gamma$ and electron temperature $kT_{\rm e}$ are essentially consistent regardless of whether the NICER data are present in spectral fitting, and hence for Insight-HXMT data without contemporary NICER data only the spectral parameters of photon index and electron temperature are considered well constrained. 
Because the minimum energy of Insight-HXMT LE can only go down to 2 keV, the Insight-HXMT data can not well constrain the parameters of the disk, resulting in a large difference from the joint fit. The results of joint fitting are more reliable for the disk parameters.

As shown in Table \ref{joint} for parameters derived from the joint fittings, the coverage factor increases gradually from MJD 58175.85 to MJD 58300.71 and decreases again at MJD 58308.99. A similar evolution trend is hinted at as well for the significance level of hard X-ray shortage $\sigma$ and electron temperature $kT_{\rm e}$.
We calculate the correlation coefficients and significance for deficit significance and the coverage factor and $kT_{\rm e}$, which were 0.89, 1.6$\sigma$ and 0.92, 1.8$\sigma$, respectively, which suggests the existence of correlation but with low significance. 
For the time-resolved spectra of the bursts, we also investigate the correlation between the average $f_{\rm a}$ of each burst and the deficit significance, and find no obvious correlation. 

With joint diagnosis from other spectral parameters e.g. photon index and flux (40--70 keV), we can see that 4U 1636--536 evolved in these four joint observations from soft to hard and then to soft.
The possible correlation of the hard X-ray shortage with the parameters of spectral index, corona temperature, and flux in 40--70 keV are investigated. As shown in Figure \ref{corr}, for a single burst all correlations turn out to be relatively weak. 
To improve the statistics, as shown in Figure \ref{CCD}, we classify the bursts in the hard state into four groups according to their positions on the CCD and merge them separately. 
Then it can be seen from the right panel of Figure \ref{corr} that the shortage significance increases with the electron temperature $kT_{\rm e}$, and has an obvious anti-correlation with the spectral index.

\subsection{Evolution of deficit fraction}
\label{deficit fraction}

\begin{figure*}
\centering

\begin{minipage}[t]{0.5\linewidth}
\centering
\includegraphics[angle=0,scale=0.3]{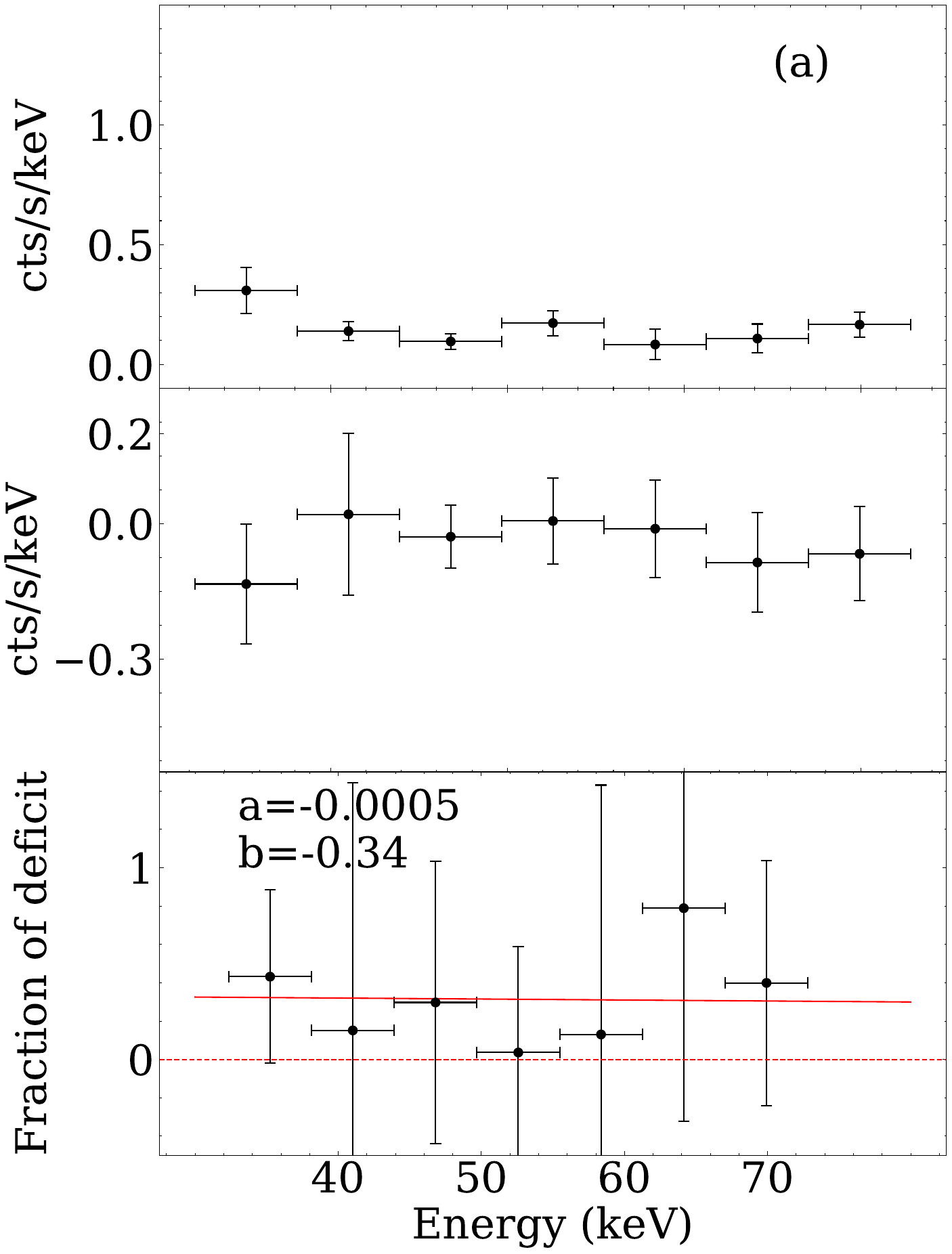}
\end{minipage}%
\hfill
\begin{minipage}[t]{0.5\linewidth}
\centering
\includegraphics[angle=0,scale=0.3]{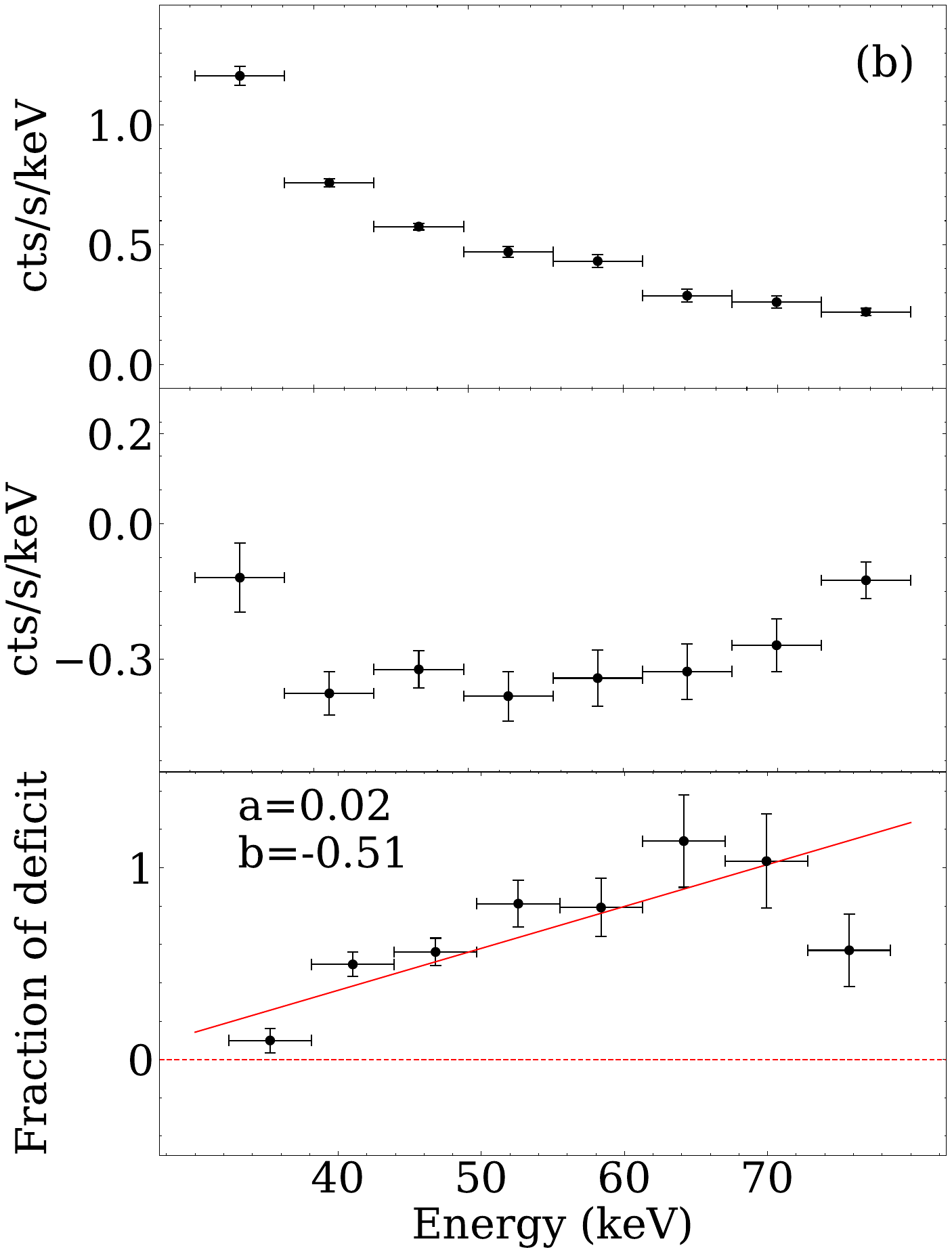}
\end{minipage}%
\hfill
\begin{minipage}[t]{0.5\linewidth}
\centering
\includegraphics[angle=0,scale=0.3]{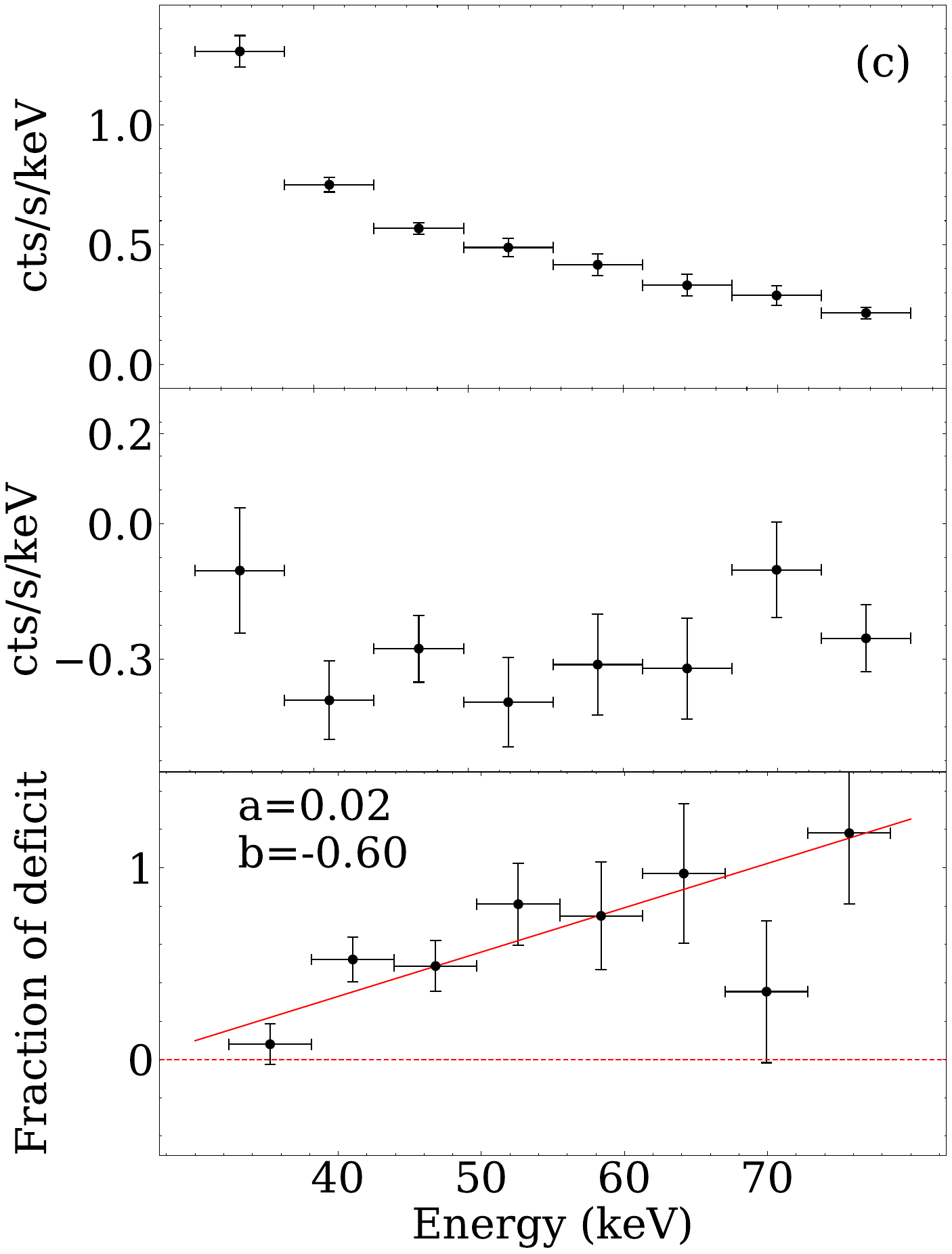}
\end{minipage}
\hfill
\begin{minipage}[t]{0.49\linewidth}
\centering
\includegraphics[angle=0,scale=0.3]{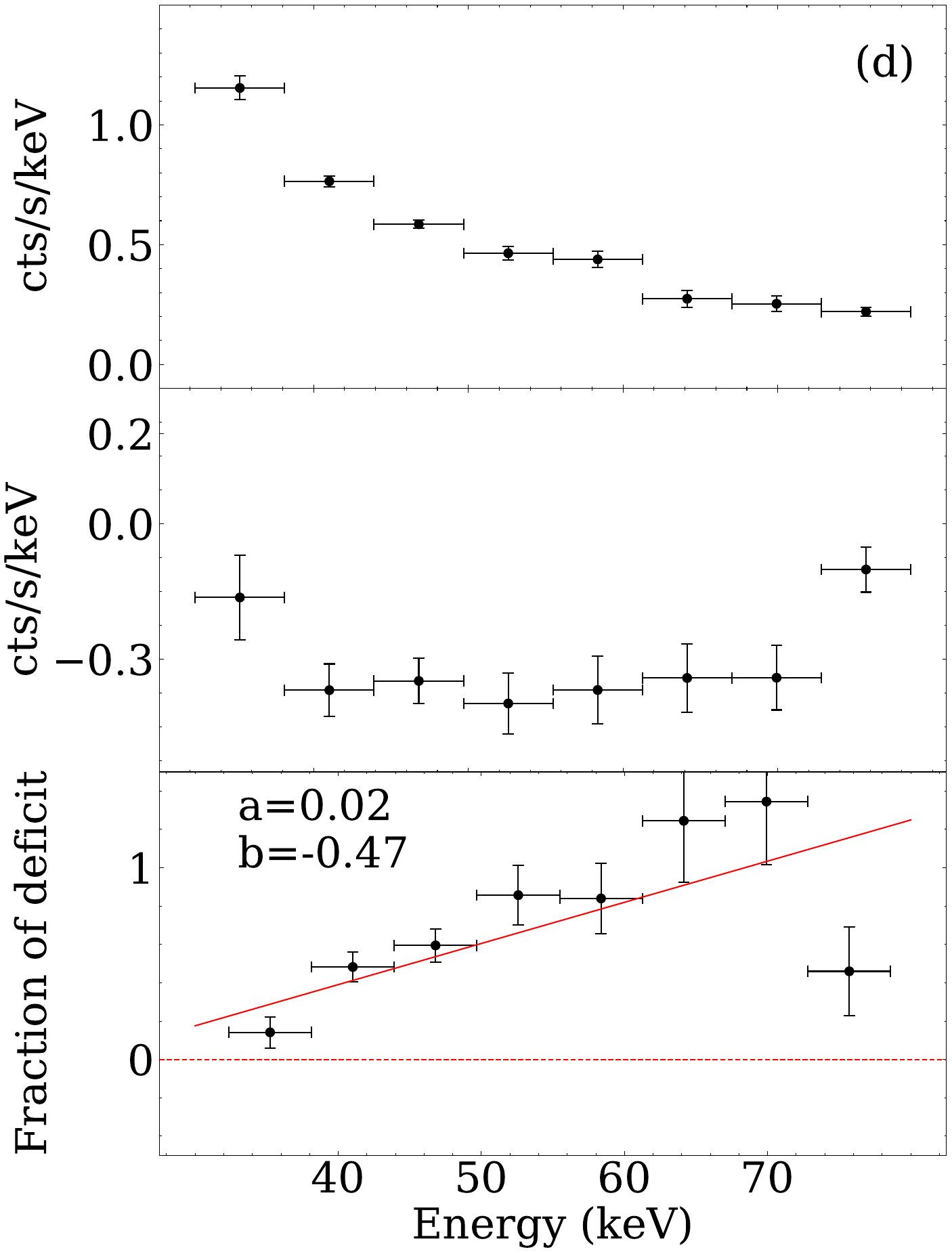}
\end{minipage}

\centering
\caption{Top panel: the spectrum of the persistent emission by HE. Middle panel: the detected spectrum of the bursts. Bottom panel: fraction of deficit VS energy during the bursts detected by HE. 
Panel (a) displays the evolution of the fraction of deficit of bursts during the soft state, panels (b), (c), and (d) show the same for the hard state, and respectively display stacked all bursts, stacked bursts with hard shortage insignificantly, and stacked bursts hard shortage with significantly during the hard state. The red solid line represents the relationship between the fraction of deficit and energy.
}
\label{burst-frc}
\end{figure*}

\begin{figure*}
\centering

\begin{minipage}[t]{0.5\linewidth}
\centering
\includegraphics[angle=0,scale=0.3]{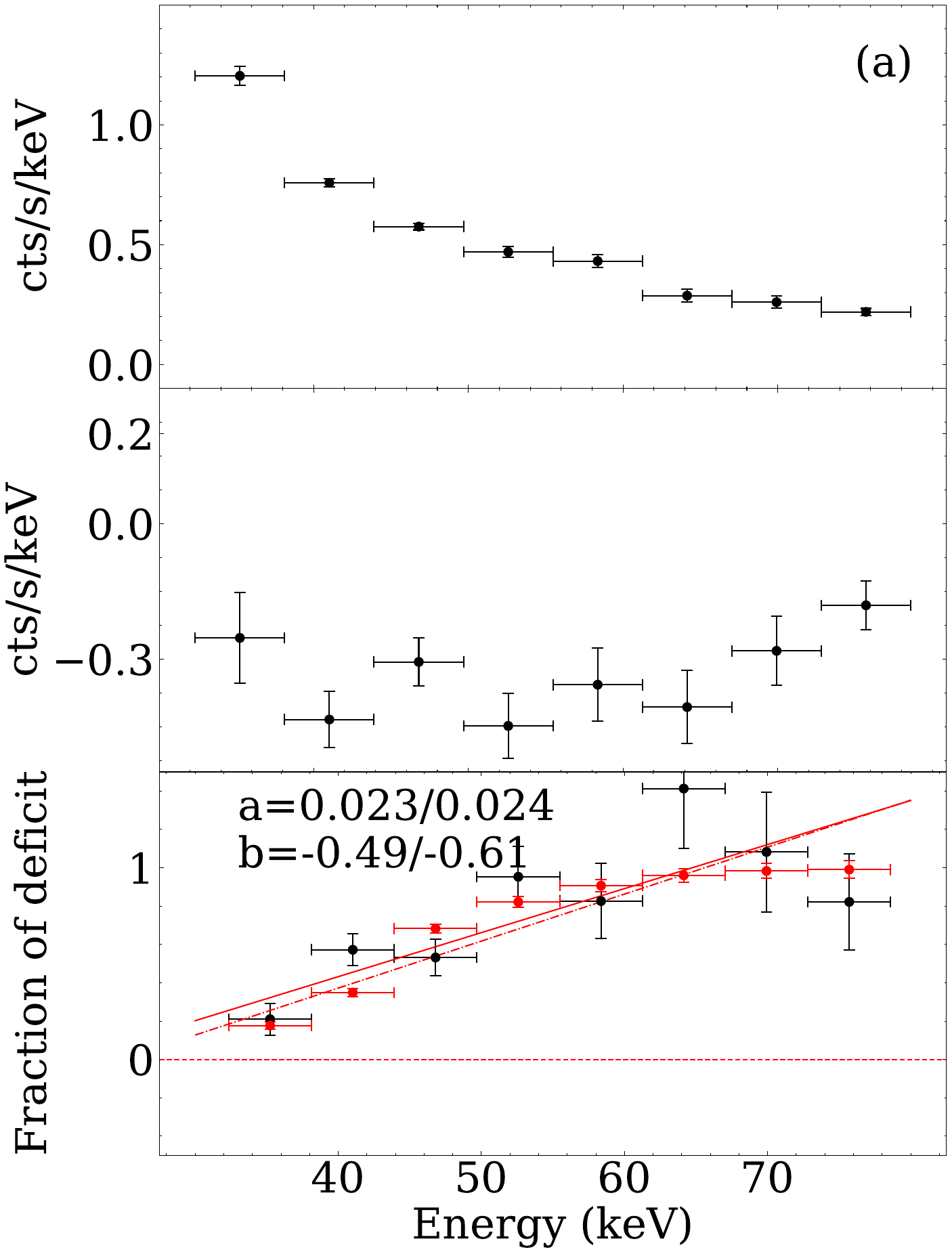}
\end{minipage}%
\hfill
\begin{minipage}[t]{0.5\linewidth}
\centering
\includegraphics[angle=0,scale=0.3]{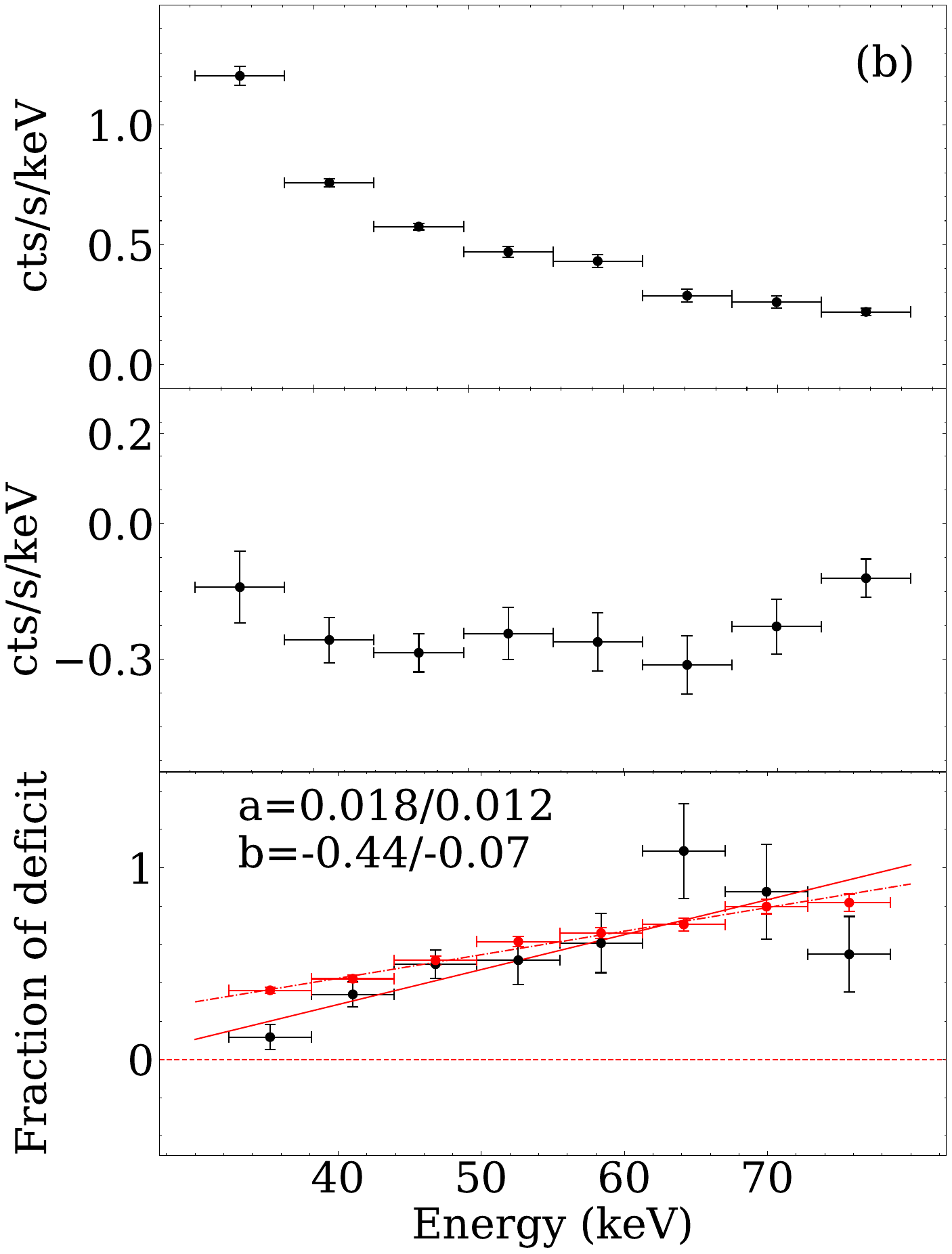}
\end{minipage}%
\hfill
\begin{minipage}[t]{0.5\linewidth}
\centering
\includegraphics[angle=0,scale=0.3]{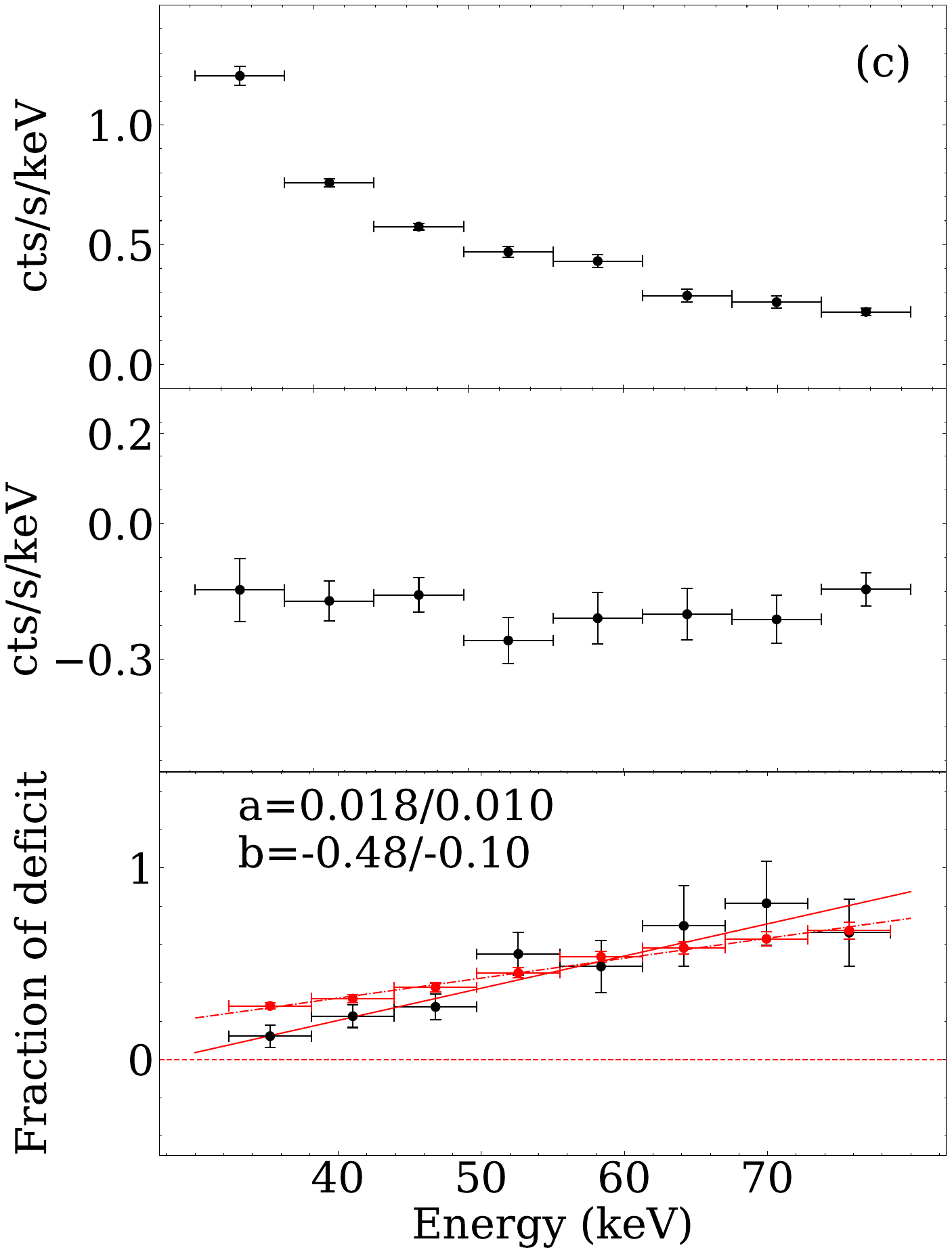}
\end{minipage}
\hfill
\begin{minipage}[t]{0.49\linewidth}
\centering
\includegraphics[angle=0,scale=0.3]{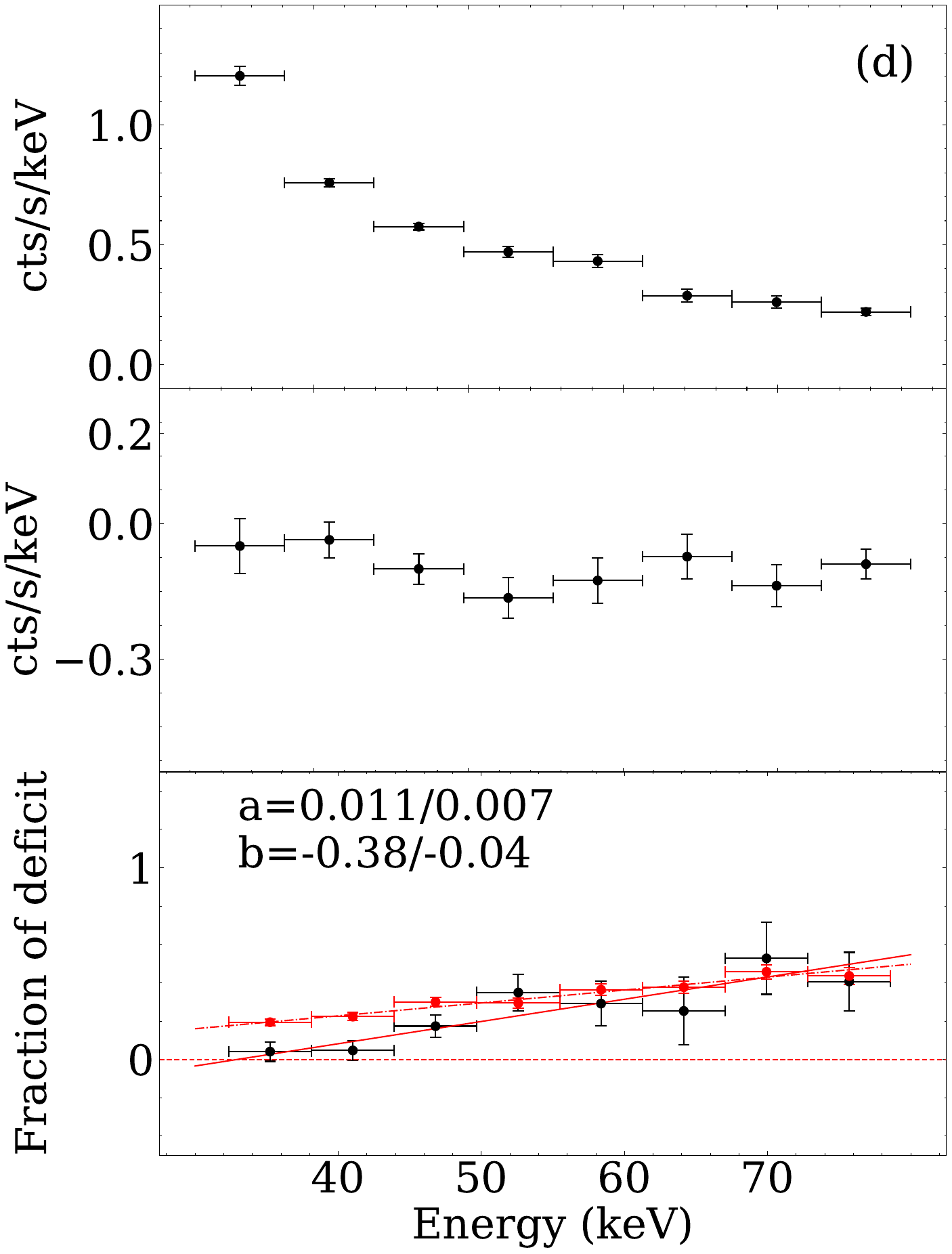}
\end{minipage}

\centering
\caption{Top panel: the spectrum of the persistent emission by HE. Middle panel: the detected spectrum of the bursts. Bottom panel: fraction of deficit VS energy during the bursts detected by HE. Panel (a) displays the peak (2--12 s) after the bursts merge during the hard state, while panels (b), (c), and (d) respectively illustrate the decay of the bursts with time intervals of 12--32 s, 32--62 s, and 62--112 s. The red points in all panels represent the simulated evolution of the deficit fraction of a corona, starting with an initial temperature of 19.8 keV and gradually being cooled down to different temperatures. The solid red line represents the measured fraction of deficit and energy, while the red dash-dot line represents the simulated fraction of deficit and energy.}
\label{frc-evo}
\end{figure*}

\begin{figure*}
\centering

\begin{minipage}[t]{0.5\linewidth}
\centering
\includegraphics[angle=0,scale=0.073]{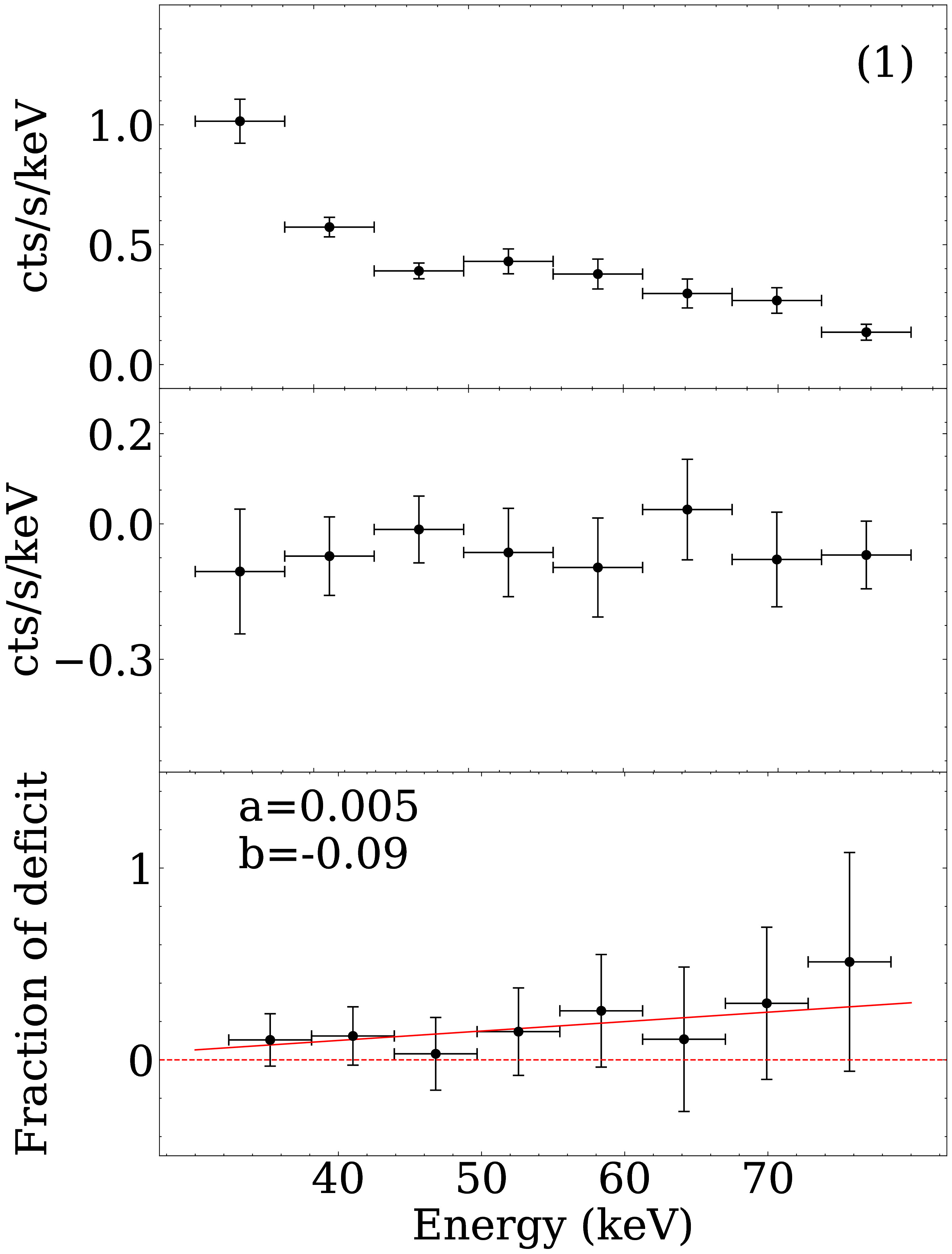}
\end{minipage}%
\hfill
\begin{minipage}[t]{0.5\linewidth}
\centering
\includegraphics[angle=0,scale=0.073]{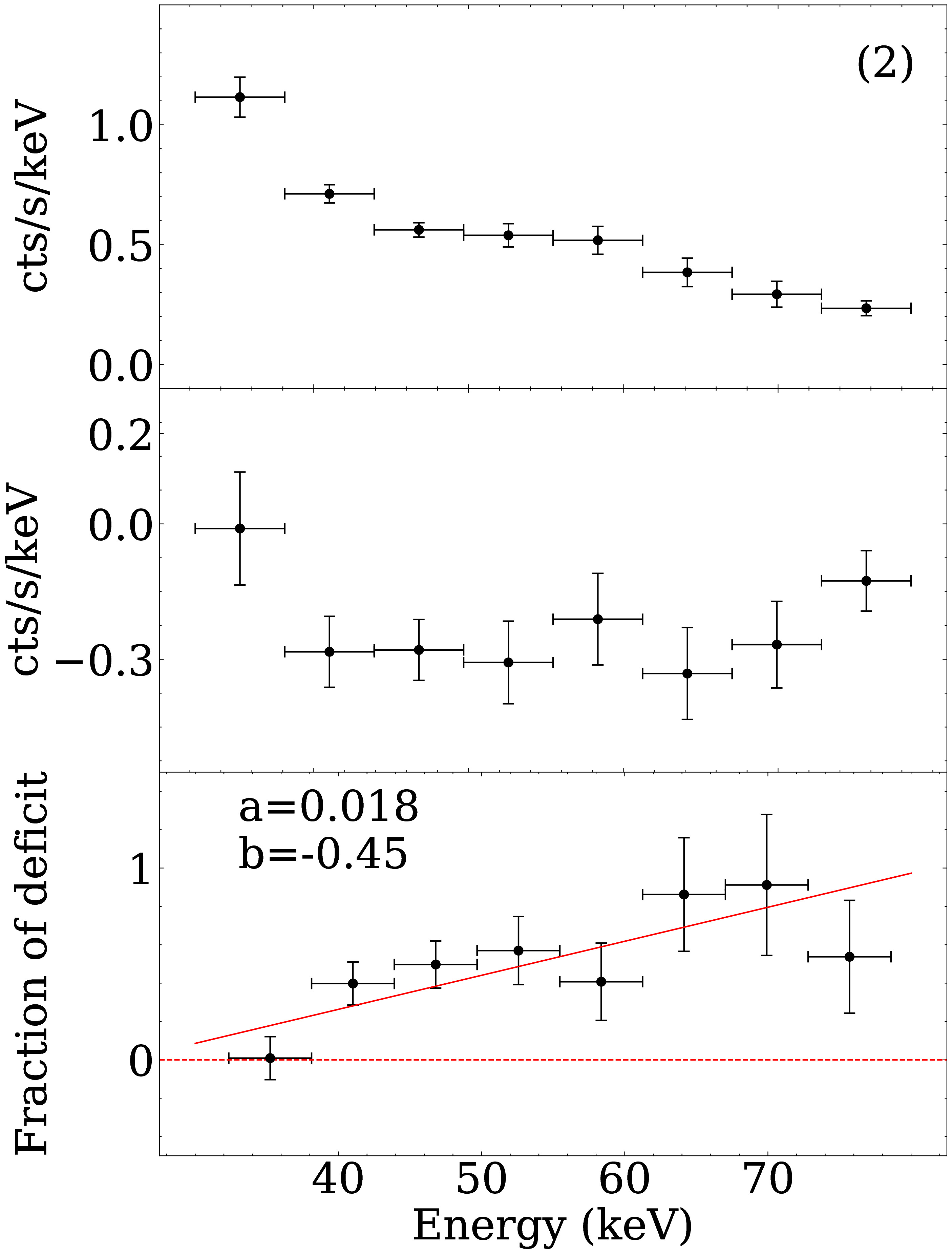}
\end{minipage}%
\hfill
\begin{minipage}[t]{0.5\linewidth}
\centering
\includegraphics[angle=0,scale=0.073]{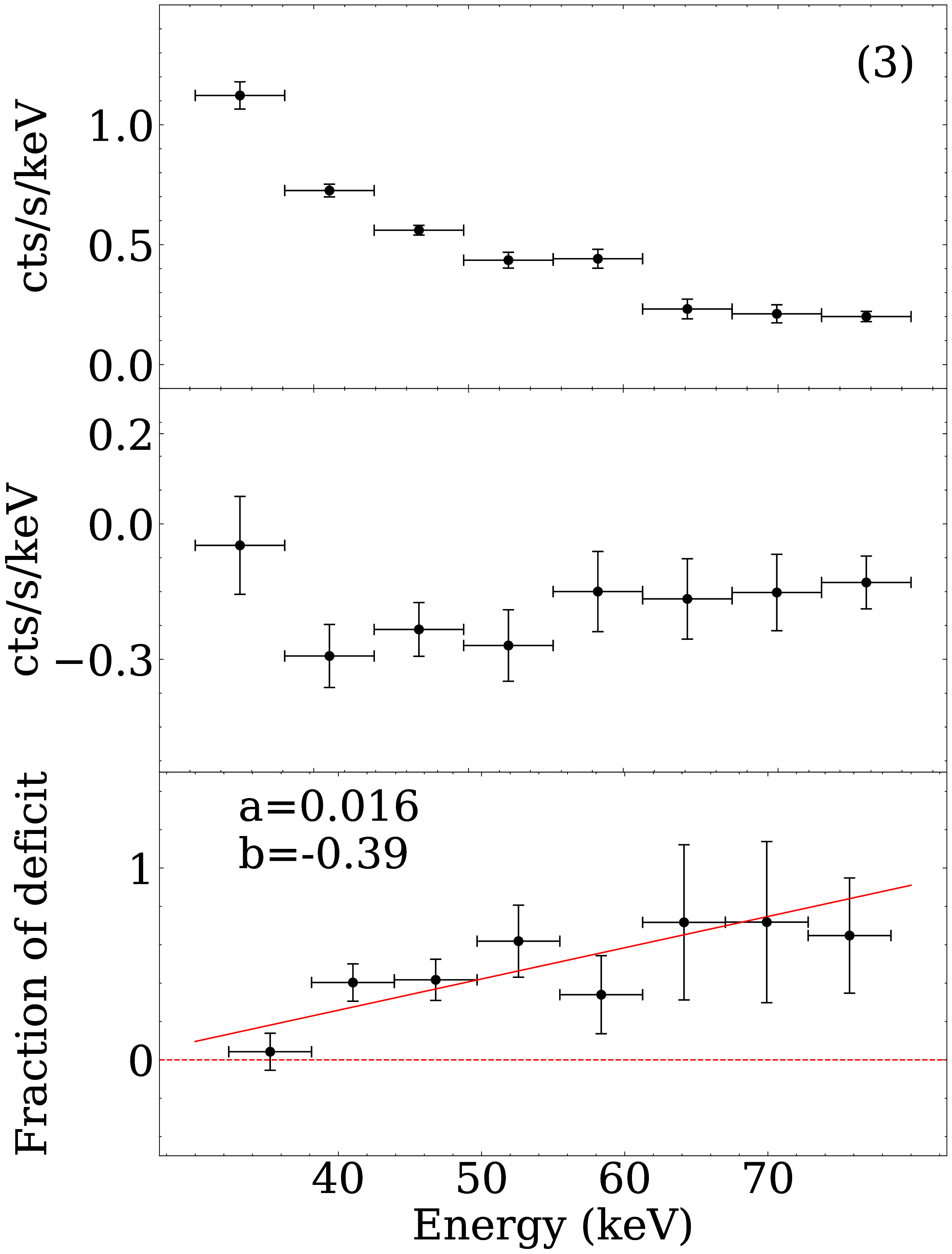}
\end{minipage}
\hfill
\begin{minipage}[t]{0.49\linewidth}
\centering
\includegraphics[angle=0,scale=0.073]{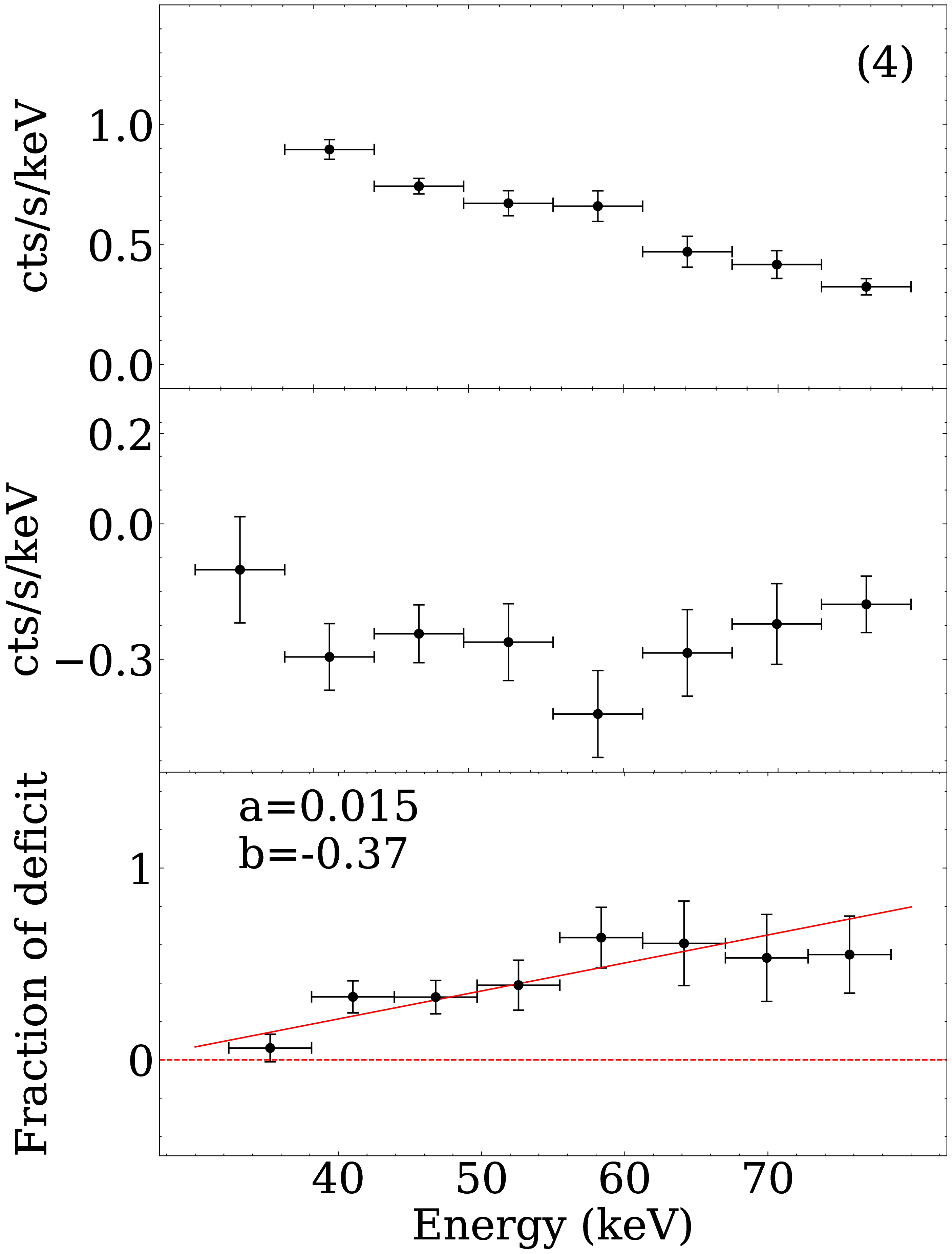}
\end{minipage}

\centering
\caption{Top panel: the spectrum of the persistent emission by HE. Middle panel: the detected spectrum of the bursts. Bottom panel: fraction of deficit  VS energy during the bursts detected by HE. Panel (1), (2), (3) and (4) correspond to the stacking of the bursts in the four regions 1, 2, 3, and 4 of Figure \ref{CCD}, respectively.  The red solid line represents the relationship between the fraction of deficit and energy. }
\label{group}
\end{figure*}

As shown in Figure \ref{lc}, some bursts exhibit a significant shortage in the light curves of HE during the bursts.
To further study the evolution of the shortage with energy, we use the tool {\tt{addspec}}\footnote[1]{{https://heasarc.gsfc.nasa.gov/ftools/caldb/help/addspec.txt}} of {\tt{HEASOFT}} to stack the spectra.
For each burst,  we extract the spectrum based on its rise time and decay time (shown in Table \ref{obs}).
Specifically, the spectra of bursts in the soft state, the hard state with low shortage significance (less than 3 $\sigma$), the hard state with high shortage significance (greater than 3 $\sigma$), and the overall hard state are combined separately. All the pre-burst spectra with an exposure time of 50 seconds before each burst are also stacked.

As shown in Figure \ref{burst-frc}, the middle panel is the spectrum of the burst after subtraction of the pre-burst persistent emission as the background. The top panel is the net spectrum of the persistent emission. We define their ratio as the fraction of the deficit (bottom panel).
We fit the relationship between the fraction of deficit and the energy with a linear function Y = aX + b.
We find that when the source is in the soft state, the fraction of the deficit is essentially zero. However, in the hard state, we find that the fraction of deficit increases with energy. This trend is observed in the spectra obtained by combining all bursts in the hard state, as well as in the spectra obtained by combining bursts according to the significance of the hard X-ray shortage. Specifically, the trend in the fraction of deficit is consistent between the bursts with low shortage significance and those with high shortage significance. 
However, bursts with significant hard X-ray shortage exhibit more dramatic evolution and larger values of the fraction of deficit compared to bursts with low shortage significance.

To investigate the evolution of the fraction of deficit during bursts, we extract the peak with a time interval of 2--12 s and the decay with time intervals of 12--32 s, 32--62 s, and 62--112 s for all bursts in the hard state. We then stack the spectra to estimate the fraction of deficit.
As shown in Figure \ref{frc-evo}, as bursts decay, the fraction of deficit also decreases. 
Similar to the energy dependence of the deficit fraction, the deficit fraction increases with energy and reaches around unity at energies around 60--70 keV and then tends to turn over. However, in 62--112 s, the fraction of deficit in the high energy band drops from 1 to about 0.3, and no trend of turnover is visible at higher energies. 
We obtain the electron temperatures of the corona by fitting the spectra for these four-time intervals, which are 4.6 keV, 7.5 keV, 12.3 keV and 16.7 keV.
We assumed that the cooling of the corona is only affected by the electron temperatures, and we obtain the evolution of the fraction of deficit using a simulation in which we only consider that the pre-burst electron temperatures of 19.8 keV are cooled down to their corresponding temperatures, respectively (red points in Figure \ref{frc-evo}). We find that the results of the simulations are in general agreement with our estimates. However, we do not find the trend of turnover at around 60-70 keV through simulations if considering only the different temperatures of the corona. 
The energy dependence of the deficit fraction is also investigated for the four groups of bursts as denoted in Figure \ref{CCD}. We find that the overall positive correlation between the energy and deficit fraction holds for groups of bursts embedded in the hard states, but with a weaker trend of turnover at high energies  (see Figure \ref{group}).

\section{Discussion and Conclusion}
\label{dis}

We have carried out detailed analyses of the burst influences on the accretion environment by taking a sample of 20 bursts detected by Insight-HXMT from 4U 1636--536. All the bursts in the hard state show hard X-ray shortage but none in the soft state. Hard X-ray shortage is found at roughly 7 $\sigma$ from one burst, serving as the most significant detection for hard X-ray shortage observed so far from a single burst. The significance of the hard X-ray shortage shows a strong anti-correlation with the spectral index of the persistent emission. The deficit fraction is found to increase with energy, suggesting either a more efficient cooling of the corona at higher energies or the entire corona is cooled down to a lower temperature. 

Since the first hint of about 2 $\sigma$ for hard X-ray shortage induced by type-I bursts was discovered \citep{2003Maccarone}, hard shortage has been found in a series of X-ray binaries, mostly via stacking a large number of bursts. However, thanks to the high energy band and large effective area of Insight-HXMT HE, hard shortage from a single burst was first detected at 6.2 $\sigma$ from 4U 1636--536 in the 40--70 keV energy range. We have analyzed the 20 thermonuclear bursts (2 in the soft state and 18 in the hard state) of 4U 1636--536 observed by Insight-HXMT and NICER from 2018 to 2022. As shown in Table \ref{obs}, hard X-ray shortage is only detected for bursts in the hard state but not in the soft state. One burst was detected with a hard X-ray shortage as significant as 7 $\sigma$, which is higher than any of the previous reports from a single burst.  A sum-up of all the bursts in the hard state results in the detection of hard X-ray shortage at a significance of 15.6 $\sigma$, higher than that of 3 $\sigma$ derived previously from stacking 114 bursts observed by RXTE \citep{2013Ji}

It is generally believed that type-I bursts can serve as a probe to study the accretion environment in LMXBRs. Apart from the increase of the persistent emission of soft X-rays during the burst, which is most likely related to the PR-dragging effect induced by the burst, the hard X-rays of the persistent emission show up with shortage, which is usually understood that the surrounding hot corona is cooled down by the soft X-ray shower of the burst. So it is not of surprise that the hard X-ray shortage is not detectable in the soft state when the source has weak hard X-ray emission, probably due to either a compact corona that exposes a very small solid angle to the neutron star, such that the corona can not effectively be cooled by the soft X-ray showers of the type-I bursts from the neutron star surface.
While in the hard state, the source is dominated by the non-thermal emission from the corona, the burst-induced hard X-ray shortage provides a chance for investigating how the corona evolves during the outburst. With a sample of 18 bursts from 4U 1636–-536, the significance of the hard X-ray shortage is found to have a strong anti-correlation with the spectral index of the persistent emission. As shown in Figure \ref{corr}, the significance increases when the spectral index decreases from 1.8 to 1.65. The non-thermal spectral index is related to the temperature and the optical depth of the corona in the form of 
$\Gamma=-\frac{1}{2}+\sqrt{\frac{9}{4}+\frac{1}{\theta \tau\left(1+\frac{\tau}{3}\right)}}$, where $\theta$=$kT_{\rm e}$/$m_{\rm e}c^2$ and $m_{\rm e}c^2$=511 keV is the electron rest mass \citep{1987Lightman}, and the optical depth $\tau$ can denote the geometric size of the corona. By taking an average of the corona temperature of around 20 keV, as shown in Figure \ref{corr}, a change of spectral index from 1.8 to 1.65 would require an increase of the corona temperature of 6.1 keV, which is slightly beyond a variation range of roughly 5 keV as shown in  Figure \ref{corr} when the bursts are classified into 4 groups. 
This in turn may suggest the need for an increment of the coronal optical depth by enlarging the size of the corona with the same density. The observed anti-correlation between the significance in shortage and the spectral index can therefore be understood, at least as one possibility, in a scenario that an enlarged corona would have a larger solid angle exposed to the neutron star and thus more efficient cooling effect by intercepting more soft X-ray photons from a burst. 
We also note that, as shown in Figure \ref{group}, the deficit fraction of the fourth group does not reach 100\%, indicating that the corona is not fully cooled, although in this group the bursts have relatively significant shortage in hard X-rays, and as well the persistent emission has a relatively hard energy spectrum. This may suggest that at the initial hard state, the corona should be further away from the neutron star if the corona size remains little change during the evolution of the hard state.

As for the burst influence on the persistent spectrum, \cite{2017Kajava}  analyzed the spectrum by stacking 123 X-ray bursts in the hard state of 4U 1728--34 using data from the \textit{INTEGRAL} JEM-X and IBIS/ISGRI instruments. They found that the emission above 40 keV decreased by a factor of approximately three during the burst with respect to the persistent emission level. We calculate the fraction of deficit for the spectrum in the counting space, as given in their paper. We find that the fraction of deficit is around $-0.3^{+0.22}_{-0.21}$ at 35 keV, and then increase to around $0.69^{+0.25}_{-0.17}$ and $0.63^{+0.35}_{-0.21}$ at 45 keV and 65 keV, respectively. These values barely show a trend of an increase in the deficit fraction with energy. A clear energy dependence of the fraction of deficit was reported from \cite{2022Chen} in a millisecond pulsar system MAXI 1816--195, where the fraction of deficit increases with energy and turns over above 70 keV. Such a turnover is interpreted as the contribution of hard X-rays from the emission region of the magnetic pole, which is less influenced by the burst shower. Here we obtain a similar result that the deficit fraction increases with energies (Figure \ref{burst-frc}).
Specifically, around the burst peak, we find the deficit fraction starts to turn over at energies above 65 keV but keeps increasing toward higher energies when the bursts decay to low flux levels (Figure \ref{frc-evo}). An overall increase of the deficit fraction with energies can be accounted for by introducing a cooler corona. For example, by having a cooled corona with a temperature of 19.8 keV, as shown by the red points in Figure \ref{frc-evo}, a deficit feature of increase with energy and flattening at high energies can be well established. However, the deficit fraction can be observed to reach almost 100$\%$ and then tend to turn over at energies above 65 keV during the burst peak period (see Figure \ref{frc-evo}). Such a feature may be accounted for if the hard X-rays at higher energies are produced in a flattened region that has a relatively smaller solid angle with respect to the neutron star and hence a decreasing cooling efficiency. In such a case the turnover would be expected to turnover gradually and eventually disappear once the corona temperature starts to recover in the decay phase of the bursts, as shown in Figure \ref{frc-evo}.

In summary, the broad energy coverage and large detection area at hard X-rays characterize Insight-HXMT as a proper mission for investigating the burst influences on the accretion environment. Accordingly, a sample of 20 bursts from 4U 1636--536 observed by Insight-HXMT provides us some insights into the possible evolution of the corona during the outburst. Apart from the usual speculations that the corona can be cooled by the burst shower, we find that the energy dependence of the deficit fraction may indicate a change of the corona properties along with the outburst.

\begin{acknowledgements}
This work is supported by the National Key R\&D Program of China (2021YFA0718500), the National Natural Science Foundation of China under grants U1838201, U1838202,  U2038101, U1938103,12173103, U1938107, 12273030, 12333007 and 12027803.
This work made use of data and software from the Insight-HXMT mission, a project funded by China National Space Administration (CNSA) and the Chinese Academy of Sciences (CAS). This work was partially supported by the International Partnership Program of CAS (Grant No.113111KYSB20190020).
This research has made use of software provided by of data obtained from the High Energy Astrophysics Science Archive Research Center (HEASARC), provided by NASA’s Goddard Space Flight Center.
\end{acknowledgements}


\bibliographystyle{aa}
\bibliography{ref}

\end{document}